\documentclass[a4paper,twocolumn,10pt,unpublished]{quantumarticle}
\pdfoutput=1
\usepackage[utf8]{inputenc}
\usepackage[english]{babel}
\usepackage[T1]{fontenc}
\usepackage{amsmath}
\usepackage{xcolor} 

\usepackage{amssymb}  
\usepackage{physics}  
\usepackage[normalem]{ulem}
\usepackage{graphicx}
\usepackage{subcaption}
\usepackage{csquotes}
\usepackage{tabularx}
\usepackage{float}
\usepackage{fixme}
\usepackage{tikz}
\usetikzlibrary{positioning}

\usepackage[numbers,sort&compress]{natbib}
\usepackage[colorlinks=true, allcolors=blue]{hyperref}

\begin{document}

\title{Differentiable Digital Twins for Optimization, Calibration, and Characterization of Neutral-Atom- and Ion-based Quantum Computers}

\author{Marco Rossignolo}
\affiliation{Qruise GmbH, 66113 Saarbrücken, Germany}
\email{rossignolomarco@gmail.com}
\orcid{}
\author{Lorena C. Bianchet}
\affiliation{Qruise GmbH, 66113 Saarbrücken, Germany}
\orcid{0000-0003-2564-231X}
\author{Anurag Saha Roy}
\affiliation{Qruise GmbH, 66113 Saarbrücken, Germany}
\orcid{0000-0002-3269-476X}
\author{Shai Machnes}
\orcid{}
\affiliation{Qruise GmbH, 66113 Saarbrücken, Germany}

\newpage
\maketitle
\pagenumbering{arabic}
\setcounter{page}{1}

\begin{abstract}
Quantum technologies based on individually controlled atoms are among the most promising platforms for scalable quantum computing, owing to their favorable coherence properties and controllability. Nonetheless, achieving reliable and high-fidelity gate operations remains challenging due to the sensitivity of these systems to noise and hardware imperfections. In this work, we introduce a comprehensive software toolset designed to model atomic quantum systems with high precision and simulate qubit gate dynamics. By integrating quantum optimal control and benchmarking techniques with automatic differentiation for efficient and accurate gradient-based optimization, the toolset enables significant improvements in gate performance. We demonstrate its capabilities through detailed case studies involving both neutral atoms and trapped ions, highlighting its effectiveness in enhancing quantum gate fidelities, and providing valuable insights for experimental implementation.
\end{abstract}

\section{Introduction}
Over the past decade, quantum technologies have evolved from primarily academic research into a rapidly developing field with commercial, security, and industrial applications. Neutral atoms \cite{SaffmanNationalScienceRev2018, Henriet2020quantumcomputing, BriegelJMO2000, Cirac2000Nature} and trapped ions \cite{Kielpinski2002, PRXQuantum.2.020343, KielpinskiNature2002, BruzewiczAPR2019} are promising platforms for large-scale fault-tolerant quantum computing because their qubits are intrinsically identical and combine precise control, strong interactions, flexible connectivity, and long coherence times.\\
Realizing this potential requires models that capture relevant device physics and control hardware, together with optimization and calibration methods that connect those models to experiment \cite{Jaschke2024, lu2025neuralquantumdigitaltwins, IEEEDigitalTwins2022, vanFrank2016, ScienceAdvances2025, PhysRevA.79.012312}. Here we present \texttt{qruise-toolset}, a differentiable-digital-twin framework building on established optimization and characterization methods \cite{PhysRevApplied.15.034080, rossignolo2023quocs, heck2018remote}. Although the examples in this work concern atomic systems, the framework is applicable to other quantum hardware platforms, including superconducting and solid-state qubits \cite{werninghaus2021leakage, RoyIEEE2025, frank2017autonomous, vetter2024gate}.\\
We first describe the Hamiltonian, control-stack, simulation, optimization, benchmarking, and calibration components of the framework. We then discuss the physical modeling requirements of neutral-atom and trapped-ion processors and demonstrate the methods using recoil-free and two-qubit neutral-atom gates, analog quantum simulation, trapped-ion gates, and Bayesian calibration.

\section{Modelling in qruise-toolset}
In this section, we describe how to initialize the system Hamiltonian using symbolic representations and how time-dependent controls are incorporated, including imperfections introduced by control electronics. We outline the various quantum dynamics simulation capabilities of the  \texttt{qruise-toolset} library, such as exact state evolution and bit-string sampling via our Qiskit-compatible backend. Furthermore, we introduce both optimal and robust optimal control techniques to enhance gate performance, and conclude with benchmarking procedures for evaluating the quantum processing unit (QPU) and calibration methods.

\subsection{Hamiltonian symbolic representation}
{A general Hamiltonian, $H$, for a quantum system can be written as}
\begin{equation}
    H = H_0 + H_D(t),
\end{equation}
where $H_0$ is the time-independent part, and $H_D(t)$ is the time-dependent driving part.\\
Since we focus on qubit evolution, we model the Hamiltonian as an interaction with auxiliary degrees of freedom—collectively referred to as \emph{motional modes}—representing all system components that couple to the qubits.\\
The time-independent part can be expressed as:
\begin{equation}
    H_0 = H_{\bar{q}} \otimes \mathbb{I}_{\bar{m}} + \mathbb{I}_{\bar{q}} \otimes H_{\bar{m}} + H_{\bar{q}_c} \otimes H_{\bar{m}_c},
\end{equation}
where $H_{\bar{q}}$ is the sum of single-particle Hamiltonians for qubits in $\bar{q} = (q_1, \dots, q_n)$,
$H_{\bar{m}}$ is the sum of single-particle Hamiltonians for motional modes $\bar{m} = (m_1, \dots, m_s)$, $H_{\bar{q}_c} \otimes H_{\bar{m}_c}$ captures couplings between qubits and motional modes as well as internal mode interactions.\\ 
The time-dependent driving part is modeled as:
\begin{equation}
    H_D(t) = H_{\bar{q}_d}(t) \otimes H_{\bar{m}_d},
\end{equation}
{where $H_{\bar{q}_d}(t)$ acts on the driven qubit subspace and $H_{\bar{m}_d}$ is the accompanying operator on the motional subspace. Time dependence is typically restricted to $H_{\bar{q}_d}(t)$.}

\begin{figure}
    \centering
    \includegraphics[width=0.9\linewidth]{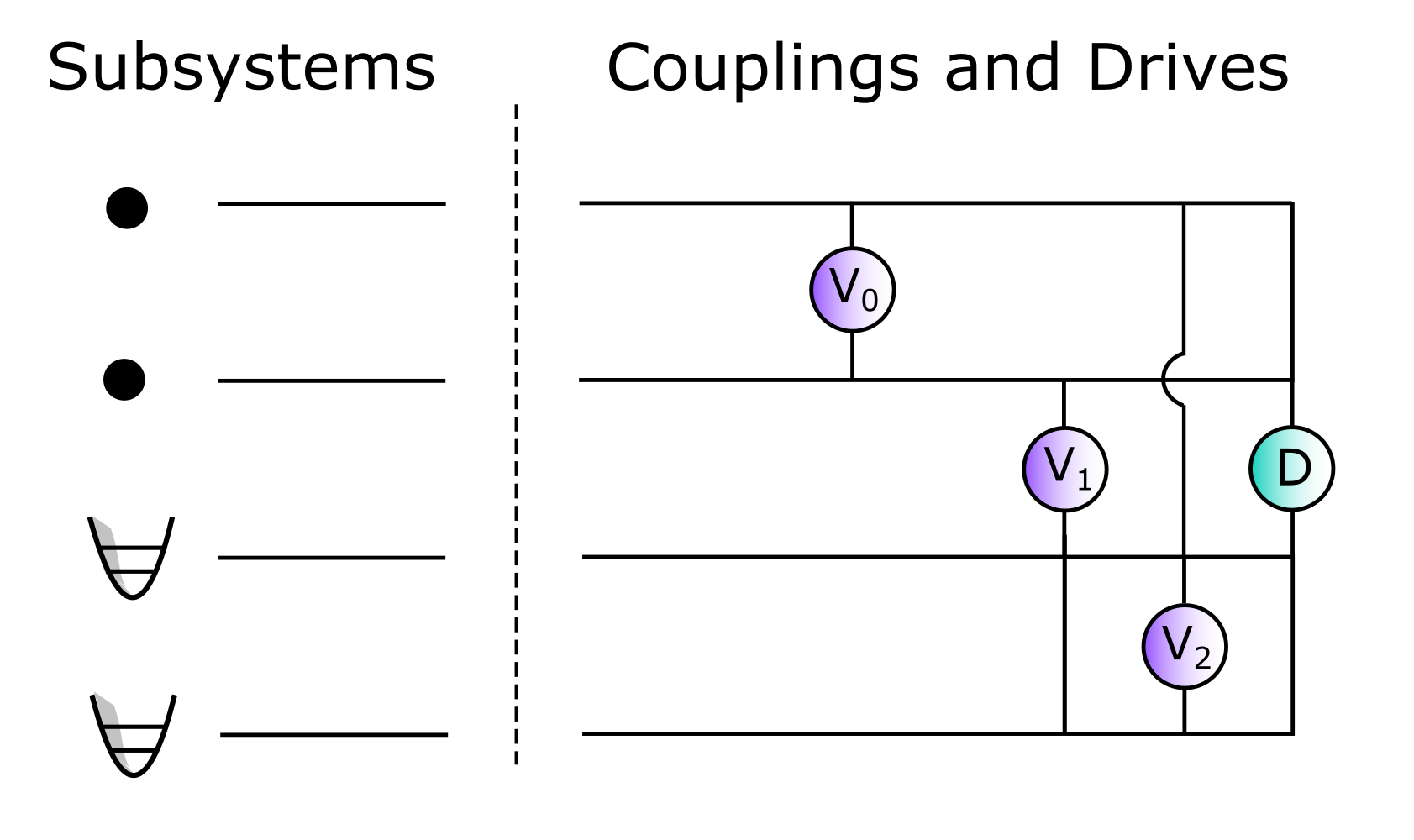}
    \caption{{Pictorial representation of the symbolic-Hamiltonian \autoref{eq:full-symbolic-hamiltonian} for a composite system. Nodes at left denote qubit (dots) and motional-mode (potential) subsystems. Coupling terms $V_i$ connect subsystems, while $D$ denotes a time-dependent drive.}}
    \label{fig:symbolic-hamiltonian}
\end{figure}

{In the \texttt{qruise-toolset} framework, a symbolic Hamiltonian is specified together with model parameters and their initial values. Its main components are subsystems, couplings, and drives.}
Each single-particle Hamiltonian, denoted as \(H_0\), is represented by a \textit{subsystem}. A subsystem is a collection of local operators that form the building blocks of the total Hamiltonian.
The interaction terms are implemented as \textit{couplings}, which link subsystems and define their mutual interactions.
The time-dependent part, $H_D(t)$, corresponds to the \textit{drives}, which incorporate time-dependent functions and symbolic representation of the drive field.
An illustration of this framework can be found in \autoref{fig:symbolic-hamiltonian}.\\
{In the framework notation, the Hamiltonian is written as}
\begin{equation}
    H = H_s + H_c + H_D(t),
    \label{eq:full-symbolic-hamiltonian}
\end{equation}
where each term is defined as follows:
\begin{equation}
    H_s = \sum_{i=1}^{N} H_{s,i}\left(\hat{\bar{O}}_i, \bar{\theta}_i\right),
\end{equation}
\begin{equation}
    H_c = \sum_{r} H_{c,r}\left(\left\{ \hat{\bar{O}}_1, \ldots, \hat{\bar{O}}_N \right\}, \bar{\theta}_r^{(c)} \right),
\end{equation}
\begin{equation}
    H_D(t) = \sum_{d} H_{D,d}\left( \left\{ \hat{\bar{O}}_1, \ldots, \hat{\bar{O}}_N \right\}, \bar{\theta}_d^{(D)}, \bar{f}_d(t), \bar{u}_d \right),
\end{equation}
where \(N\) is the total number of subsystems, \( \hat{\bar{O}}_i \) is the set of local operators of the $i^\text{th}$ subsystem, \( \bar{\theta}_i \), \( \bar{\theta}_r^{(c)} \) and \( \bar{\theta}_d^{(D)} \) are the model parameters associated with subsystem \( i \), the $r^\text{th}$ coupling term,
and the $d^\text{th}$ drive, respectively. 
The set of symbols \( \bar{f}_d(t) \) are arbitrary time-dependent modulation functions, and \( \bar{u}_d \) represents the set of symbolic control fields.
\paragraph{Qubits and Motional Modes as Subsystems}
{For a minimal system comprising one qubit and one motional mode, the Hamiltonian is}
\begin{equation}
    H = H_q \otimes \mathbb{I}_m + \mathbb{I}_q \otimes H_m,
\end{equation}
{where \( H_q \) and \( H_m \) are the single-particle Hamiltonians for the qubit and motional mode, respectively, and \( \mathbb{I}_q \) and \( \mathbb{I}_m \) are identity operators on the respective subspaces.}\\
A qubit Hamiltonian may take the form:
\begin{equation*}
    H_q = \frac{\Delta}{2} \sigma_z,
\end{equation*}
where \( \Delta \) is the qubit detuning and \( \sigma_z \) is the Pauli-Z operator. In  \texttt{qruise-toolset}, \( \Delta \) is a model parameter and \( \sigma_z \) is a local operator. The subsystem can also contain other local operators like number operators and raising or lowering operators. This allows the definition of the couplings and drives later. \\ 
A typical motional Hamiltonian is:
\begin{equation*}
    H_m = \omega_m \hat{a}^\dagger \hat{a},
\end{equation*}
where \( \omega_m \) is the mode frequency, and \( \hat{a}\) and \(\hat{a}^\dagger \) are the annihilation and creation operators of the motional mode, respectively.

\paragraph{Qubit interaction as coupling}
As a coupling example, the Van der Waals interaction between two qubits can be modelled as:
\begin{equation*}
    H_{c,q} = V_r \, n_1 n_2,
\end{equation*}
where \( V_r \) is the interaction strength and \( n_i \) is the projector (number operator) acting on the $i^\text{th}$ qubit subsystem. In  \texttt{qruise-toolset}, the interaction strength, \( V_r \), is a model parameter, and \( n_i \) is a local operator that has to be defined in the particle subsystem.

\paragraph{Drive}
A typical example of a drive is:
\begin{equation*}
    H_D(t) = \frac{\Omega(t)}{2} \left( \sigma_+ e^{i \eta (\hat{a} + \hat{a}^\dagger)} + \text{H.c.} \right),
\end{equation*}
where \( \Omega(t) \) is the time-dependent Rabi frequency defined via a symbolic control field, \( \eta \) is the Lamb-Dicke parameter that quantifies the strength of the qubit-motional coupling, and \( \sigma_+ \) is the raising operator acting on the qubit. In  \texttt{qruise-toolset}, such a drive is constructed by selecting the appropriate subsystem operators (e.g., \( \sigma_+ \), \( \hat{a} \)),
defining symbolic expressions involving exponentials or other operator combinations, and
linking the symbolic Hamiltonian term to a user-defined control signal, \( \bar{u}_d\)(e.g., \( \Omega(t) \) is the symbolic representation of the control signal \( u \)).
\paragraph{Remark}
All symbols and notations introduced in this section will be consistently used throughout the manuscript to ensure clarity and continuity.

\subsection{Control stack modelling}
\begin{figure*}
    \centering
    \includegraphics[width=0.8\textwidth]{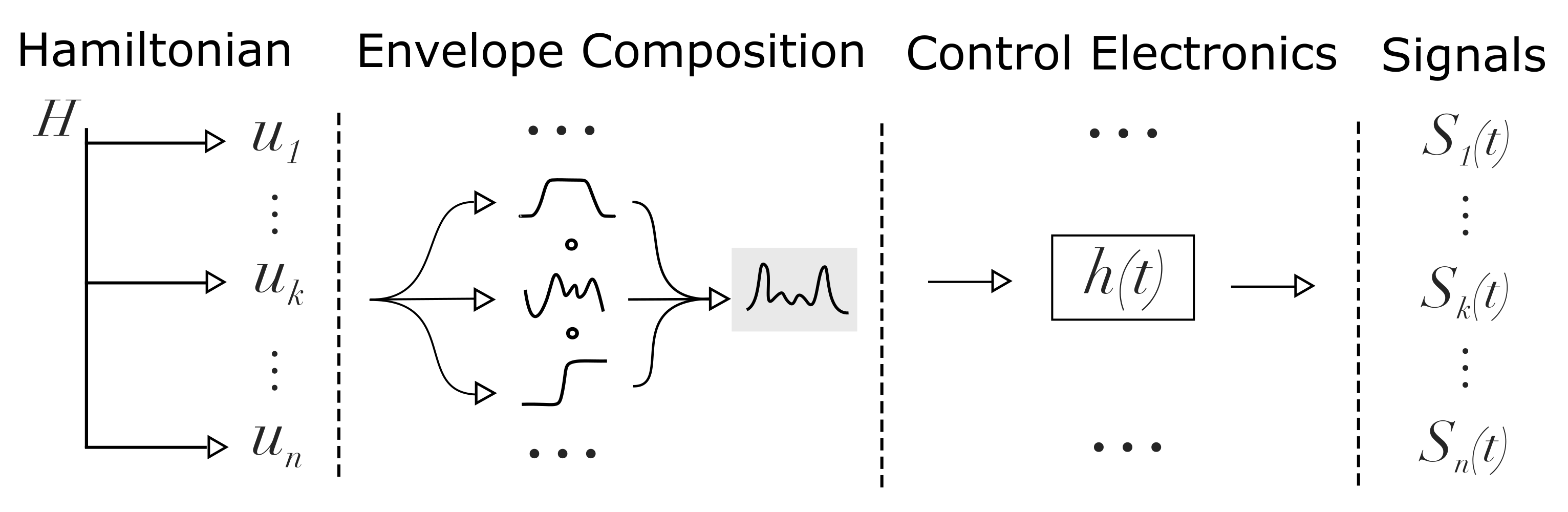}
    \caption{For each symbolic drive $u_{k}$, the user has to associate an envelope, e.g. piecewise constant, Fourier basis, sigmoid, or a composition of these. After that, the composite envelope passes through the control electronics, where the hardware imperfections change the shape through a composition of user response functions, summarized by $h(t)$ in the figure. Finally, the resulting signal $\mathcal{S}_k(t)$ is used in the system simulation.}
    \label{fig:control-stack-process}
\end{figure*}

The control stack is a layered system responsible for generating, delivering, and managing the physical control signals that manipulate and read out qubit states. To replicate the behaviour of diverse experimental setups, the qruise software provides a flexible and modular framework for modelling quantum control stacks. This includes components such as laser sources, arbitrary waveform generators (AWGs), analog-to-digital converters and local oscillators, among others. Users define the external driving fields through envelope objects, which specify the temporal profile of control signals. These envelopes can be arbitrary piecewise-defined shapes or expansions in a specific basis, such as the Fourier basis. All parameters defining the control fields are collected in the set \(\bar{\varphi}^{(e)}\).
{By incorporating the behavior of electronic components that generate and shape control signals,  \texttt{qruise-toolse} supports realistic control-layer simulations. Predefined stack models emulate standard AWG behavior, while custom components represent experiment-specific hardware. Hardware parameters form the set \(\bar{\varphi}^{(s)}\). Varying stack and envelope parameters quantifies how hardware characteristics and signal distortions affect gate performance. The resulting control pulses construct the driving Hamiltonian and enable fidelity simulations under realistic noise conditions (\autoref{fig:control-stack-process}).}

\subsection{Simulation}
{After building the Hamiltonian system and defining the control stack, the propagation methods simulate closed and open quantum dynamics (Appendix \ref{app:closed-open-quantum-systems}). Stochastic fluctuations with arbitrary power spectral densities can be injected through the noise-generation toolkit. A custom Qiskit \cite{qiskit2024} backend exposes model-based QPU measurement outcomes.}
\subsubsection{Noise Generation}
To accurately simulate realistic quantum dynamics, noise processes with arbitrary power spectral density (PSD) can be injected into the control parameters. 
One method for generating such noise traces involves constructing a real-valued, zero-mean Gaussian process \( X(t) \) with a target PSD \( S_X(f) \), using a Fourier-based synthesis approach:
\[
X(t) = \sum_{j=1}^{\infty} 2\sqrt{S_X(f_j)\Delta f} \cos(2\pi f_j t + \varphi_j),
\]
where the phases \( \varphi_j \) are independently drawn from the uniform distribution \( \mathcal{U}[0, 2\pi] \), and the frequencies are defined as \( f_j = j \Delta f \). This construction ensures that the generated time trace exhibits the correct statistical variance,
\[
\sigma_X^2 = 2 \sum_{j=1}^{\infty} S_X(f_j) \Delta f,
\]
faithfully reproducing the desired spectral characteristics of the noise process.
An efficient alternative for generating stochastic time series with a given PSD is based on the inverse fast Fourier transform (FFT). The method proposed in~\cite{timmer1995generating, welch2003use} proceeds by defining the target PSD, \( S(\omega) \), over a discrete frequency grid. For each frequency, \( \omega_i \), we generate two independent standard normal random variables to construct a complex-valued Fourier component \( f(\omega_i) \), scaled as
\[
f(\omega_i) = \sqrt{\tfrac{1}{2} S(\omega_i)} \, (a_i + i b_i),
\]
where \( a_i, b_i \sim \mathcal{N}(0, 1) \). To ensure a real-valued time trace, we enforce Hermitian symmetry in the frequency domain via \( f(-\omega_i) = f^*(\omega_i) \), and set the Nyquist frequency component to be real-valued. Applying the inverse FFT to this symmetrized frequency-domain signal yields a real-valued stochastic process \( X(t) \) with the desired spectral and statistical characteristics. This method avoids artifacts such as spurious peaks associated with discrete dominant frequencies, providing smooth and reliable noise traces.
\subsubsection{Qiskit backend}
The  \texttt{qruise-toolset} provides a custom Qiskit backend built on top of the user-defined system Hamiltonian and control stack. This backend allows seamless integration with the Qiskit ecosystem, enabling tasks such as randomized benchmarking, transpilation, and quantum volume estimation. Users can simulate the execution of Qiskit circuits and retrieve realistic measurement outcomes, bridging the gap between idealized gate-level circuits and physically informed hardware models.

\subsection{Optimal control}
{The performance of quantum operations is often reduced by noise and coupling to external degrees of freedom. In this work, quantum optimal control (QOC) shaped time-dependent controls to maximize fidelity-based objective functions. Gradient-based algorithms from the BFGS \cite{lbfgsb1997} or SGD \cite{ruder2016overviewsgd} families iteratively updated the control parameters. Robust QOC incorporated parameter uncertainties directly in the objective to identify controls that maintained high fidelity across the modeled variations. Applications of QOC and robust QOC appear in \autoref{sec:applications}.}

\subsubsection{Automatic differentiation}
The symbolic Hamiltonian, control envelopes, and propagation routines are differentiable with respect to model and control parameters. Automatic differentiation (AD) evaluates derivatives of a chosen objective function through the complete simulation without finite-difference approximations. In contrast, finite-difference gradients require separate simulations for perturbed parameter values and depend on a manually selected perturbation size: overly large steps introduce truncation error, whereas overly small steps amplify numerical noise. For a model with many control parameters, finite differences also scale linearly with the parameter count, while reverse-mode automatic differentiation evaluates the full gradient at a computational cost comparable to a small number of simulations. %
The AD engine supports both built-in and custom-defined cost functions based on fidelity metrics. Once a cost function is specified, gradient-based optimization algorithms are employed to iteratively adjust the control parameters and converge toward high fidelity shapes.
\subsection{Benchmarking and calibration}

Once QOC pulses have been designed to implement high-fidelity gate operations, experimental validation becomes crucial.  \texttt{qruise-toolset} offers a comprehensive suite of benchmarking and calibration tools to assess gate performance and iteratively refine system parameters. This section focuses on quantum process tomography and Bayesian calibration methods. Additional details on implementing randomized benchmarking and quantum volume protocols are provided in Appendix~\ref{app:benchmark-techniques}.
\subsubsection{Quantum Process Tomography}

Assuming an initially uncorrelated qubit–motional mode state, the reduced dynamics of the qubit—treated as an open quantum system—are described by a completely positive map,
\begin{equation*}
\mathcal{E}(\rho) = \sum_i A_i \rho A_i^{\dagger},
\end{equation*}
where the Kraus operators $\{A_i\}$ satisfy $\sum_i A_i^{\dagger} A_i \leq I$, with equality in the trace-preserving case. The total Hilbert space is defined as \(\mathcal{H} = \mathcal{H}_{\text{qubit}} \otimes \mathcal{H}_{\text{motion}}\). To reconstruct the map \(\mathcal{E}\), we express it in a fixed operator basis \(\{E_i\}\) for \(\mathcal{B}(\mathcal{H})\), chosen to satisfy the orthogonality condition:
\[
\operatorname{Tr}(E_i^\dagger E_j) = d\, \delta_{ij},
\]
where \(d = \dim(\mathcal{H})\). In this basis, the process can be written using the $\chi$-matrix representation:
\begin{equation}
\mathcal{E}(\rho) = \sum_{i,j} \chi_{ij} E_i \rho E_j^\dagger,
\end{equation}
where the $\chi$ matrix is positive semi-definite and contains all information about the quantum process. This description naturally arises from a joint unitary evolution followed by a partial trace over the motional degrees of freedom:
\[
\mathcal{E}(\rho_{\text{qubit}}) = \operatorname{Tr}_{\text{motion}}[U(\rho_{\text{qubit}} \otimes \rho_{\text{motion}})U^\dagger].
\]
From this, one can define effective Kraus operators acting only on the qubit subspace, allowing the dynamics to be expressed as:
\[
\mathcal{E}(\rho_{\text{qubit}}) = \sum_{i,j} \chi_{ij} E_{q,i} \rho_{\text{qubit}} E_{q,j}^\dagger,
\]
where \( \{E_{q,i}\} \) form a basis for \( \mathcal{B}(\mathcal{H}_{\text{qubit}}) \). This formulation enables accurate characterization of gate operations in the presence of entanglement or decoherence with motional modes.
\paragraph{Single qubit process tomography}
In the single qubit case, the process tomography simplifies significantly \cite{Nielsen_Chuang_2010}. Using the Hermitian operator basis
\[
\tilde{E}_0 = I,\quad \tilde{E}_1 = X,\quad \tilde{E}_2 = -iY,\quad \tilde{E}_3 = Z,
\]
the process matrix $\chi$ becomes a $4 \times 4$ matrix, with up to 12 independent parameters for trace-preserving maps. By preparing four input states ($|0\rangle$, $|+\rangle$, $|-\rangle$, and $|1\rangle$) and applying $\mathcal{E}$, one obtains the corresponding output states, which are then used to construct a block matrix. A fixed congruence transformation with
\[
\Lambda = \frac{1}{2}
\begin{bmatrix}
I & X \\
X & -I
\end{bmatrix}
\]
yields the reconstructed $\chi$ matrix. Diagonalizing $\chi$ allows extraction of the Kraus operators:
\[
A_k = \sqrt{d_k} \sum_j U_{jk} \tilde{E}_j,
\]
where $d_k$ and $U$ are eigenvalues and eigenvectors of $\chi$. For a comprehensive treatment, the reader is referred to Appendix~\ref{app:single-qubit-qpt}. 
\subsubsection{Bayesian calibration}
\label{subsubsec:BayesianCalibrationTheory}

Calibrating the classical parameters governing quantum gates—such as detunings, Rabi frequencies, and trap depths, among others—is critical for high-fidelity operation. Traditional methods, such as parameter sweeps, scale poorly and fail to account for experimental uncertainty.
Bayesian calibration offers a systematic alternative to parameter sweeps. A prior distribution encodes existing knowledge of system parameters, while the likelihood quantifies the probability of the measured data for each parameter value. Bayes' theorem combines these quantities into a posterior distribution:
\[
\textit{Posterior} \propto \textit{Likelihood} \times \textit{Prior}.
\]
This \textit{Posterior} is iteratively updated as new experimental data become available, refining the estimates while quantifying uncertainty. The result is an efficient and robust calibration loop capable of navigating complex parameter landscapes and noisy environments.
A detailed use case example is described in \autoref{subsec:BayesianCalibration}.

\section{Applications to atomic systems}
\label{sec:applications}
{This section applies the framework to neutral-atom and trapped-ion systems. We first summarize the physical effects that must be represented in atomic models, then present neutral-atom recoil-free, controlled-Z, and analog-simulation case studies, followed by trapped-ion gate and calibration examples.}

\subsection{Atomic-system modelling requirements}
\begin{figure}[!htb]
 \centering 
 \includegraphics[trim = 55 0 55 15, clip, scale=0.35]{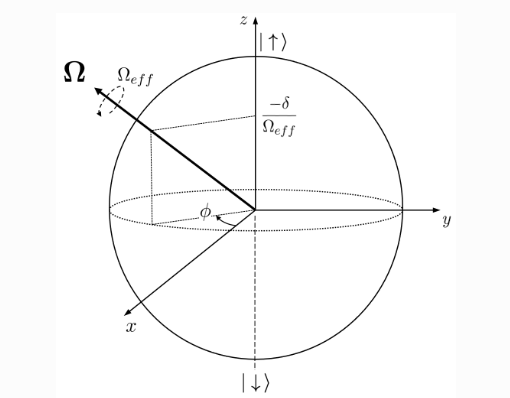} 
 \caption{{Bloch-sphere representation of a laser-driven two-level atom. The effective rotation vector is $\bar{\Omega}(t)=[\Omega(t)\cos\phi,-\Omega(t)\sin\phi,-\delta(t)]^T$, where $\Omega(t)$ is the Rabi frequency, $\delta(t)$ the detuning, and $\phi$ the drive phase.}}
 \label{fig:bloch_sphere}
\end{figure}

{Atomic quantum processors comprise individually trapped particles whose two-level electronic states encode qubits. Time-dependent electromagnetic fields, typically laser fields, coherently drive transitions and mediate gate interactions.}\\
Each control signal is tuned to address a specific atomic transition coherently. For example, targeting the transition between the ground state $\ket{g}$ and an excited state $\ket{e}$, these levels act as the $\ket{0}$ and $\ket{1}$ states of a qubit, forming a quantum spin-$\frac{1}{2}$ system.\\
{A laser pulse applied to atom $i$, with Rabi frequency $\Omega(t)$, detuning $\delta(t)$, and phase $\phi$, gives the internal Hamiltonian}
\begin{equation}
\frac{H}{\hbar} = \frac{1}{2}\bar{\Omega}(t)\cdot \bar{\sigma},
\end{equation}
with $\hbar$ being the reduced Planck constant, $\hat{\Omega}(t) = (\Omega(t) \cos (\phi), -\Omega(t) \sin (\phi), -\delta(t))^{T}$, and $\hat{\sigma} = (\sigma^x, \sigma^y, \sigma^z)^T$. In the Bloch sphere picture, this corresponds to a rotation around the vector $\bar{\Omega}(t)$ with instantaneous angular velocity $\Omega_{\text{eff}} = \sqrt{\Omega(t)^2 + \delta(t)^2}$ as shown in \autoref{fig:bloch_sphere}.
\paragraph{Effect of motional modes}
In a realistic setup, each atom is not perfectly localized but undergoes quantized motion in its trapping potential. These motional modes couple to the internal dynamics through the laser field, especially in tightly focused beams or when the wavelength of the light is comparable to the spread of the atomic wavefunction. This leads to state-dependent forces and introduces motional sidebands in the spectrum.\\
{To capture this effect, the internal Hamiltonian is extended to include motional degrees of freedom. For a single motional mode with creation and annihilation operators $a^\dagger$ and $a$, and trap frequency $\omega$, the Hamiltonian becomes}
\begin{equation}
\frac{H}{\hbar} = \frac{1}{2} \bar{\Omega}(t) \cdot \bar{\sigma} \otimes \mathbb{I} + \omega \left(a^\dagger a + \frac{1}{2} \right) + \frac{H_{\text{int}}}{\hbar},
\end{equation}
where $H_{\text{int}}$ includes the coupling between spin and motion. In the Lamb-Dicke regime, where the spatial extent of the atomic wavepacket is much smaller than the laser wavelength, the leading-order coupling is:
\begin{align*}
\frac{H_{\text{int}}}{\hbar} \approx i \eta \Omega(t) \left( a e^{-i \omega t} - a^\dagger e^{i \omega t} \right) \\
\left( \cos(\phi) \sigma_y + \sin(\phi) \sigma_x \right),
\end{align*}
with $\eta$ the Lamb-Dicke parameter. This term describes sideband transitions that can either excite or de-excite motional levels depending on the laser frequency and phase.
These couplings can degrade gate fidelity if motional excitation is not properly accounted for or suppressed. However, they can also be exploited for multi-qubit gates (e.g., in entangling schemes based on shared motional modes), motional sideband cooling, or Hamiltonian engineering.\\
Thus, driving two-level transitions in neutral atom devices is intrinsically a spin-motion coupled problem, and optimal control strategies must often take both internal and motional dynamics into account to ensure high-fidelity quantum operations.

\paragraph{Gate errors induced by motional effects}
Quantum gates in cold atoms quantum processors are inevitably affected by noise. While many sources of error are well understood—such as imperfections in state preparation and measurement (SPAM), laser inhomogeneities, and dephasing noise,one particularly insidious source arises from the motion of the atoms themselves. Even when atoms are cooled close to their motional ground states, residual thermal excitations and photon recoil can lead to entanglement between internal and motional states during gate operations, ultimately reducing gate fidelity.\\
To realistically simulate these quantum gates, we must model both conventional noise sources and motion-induced errors. SPAM errors include imperfect preparation of the internal qubit state and misidentification during state detection. For example, an atom prepared in one internal state may undergo spontaneous decay, off-resonant optical pumping, or a state-changing collision before detection and consequently be assigned to the wrong qubit state. Conversely, atom loss or imperfect fluorescence collection can cause an atom prepared in the other qubit state to be misidentified. The relative importance of these mechanisms depends on the atomic species, qubit encoding, and detection protocol.
Additionally, the finite waist of the addressing laser means that atoms located off-center experience reduced Rabi frequencies, introducing spatial inhomogeneities in the effective gate amplitude across the array. Dephasing noise arises from random interactions with the environment, which can be modelled as probabilistic $Z$-rotations.
\paragraph{Thermal effects}
Thermal motion introduces Doppler shifts in the laser frequency seen by the atoms, leading to a time-dependent detuning and potentially damping the atom population. At typical experimental temperatures of tens of $\mu\mathrm{K}$, this effect can cause systematic under-rotations or detuning drifts. Its frequency-dependent modulation can also interfere with the motional sidebands.
However, it is the motion of the atoms during the laser-driven gate that introduces fundamentally different and often harder-to-suppress errors. Even after laser cooling, atoms occupy a thermal distribution of vibrational states in their trapping potentials. We describe this distribution by a thermal density matrix, characterized by an average phonon number. While high-fidelity cooling can prepare atoms with ground-state population probabilities close to 1, a more conservative estimate of 90\% is appropriate in realistic gate sequences where photon recoil from repeated pulses gradually heats the atoms. This assumption ensures that optimized control pulses are robust not only at initialization, but also under heating accumulated across long gate sequences.
\paragraph{Different regimes}
Several regimes illustrate how motion affects gate fidelity. In the so-called \emph{frozen motion} regime, the Rabi frequency, $\Omega$, is much larger than the trap frequency, $\omega$, meaning the atom undergoes Rabi oscillations much faster than its mechanical motion. In this case, motional sidebands are only weakly excited. The probability to excite a phonon during a $\pi$ pulse scales as $\left( \pi \eta \omega / \Omega \right)^2$, where $\eta$ is the Lamb-Dicke parameter. This regime is often accessible when using tightly focused addressing beams along the trap axis, enabling high-speed gates with minimal motion-induced error.\\
In contrast, the \emph{resolved-sideband} regime corresponds to the opposite timescale limit, $\Omega \ll \omega$. In this regime, the spectral width of a carrier-resonant drive is much smaller than the spacing between motional sidebands. Carrier and sideband transitions can therefore be addressed selectively, and off-resonant excitation of motional sidebands is suppressed. Unlike the frozen-motion regime, this suppression arises from spectral resolution rather than from a gate duration short compared with the motional period. Conversely, tuning the drive to a sideband enables selective phonon-changing transitions.\\
From a practical standpoint, we must design quantum gates that remain robust despite thermal fluctuations and experimental imperfections. This includes accounting for atoms starting in mixed motional states, as well as trap inhomogeneities across the array of tweezers.

\paragraph{Entanglement}
Aside from the impact on fidelity, it's important to quantify how much entanglement is created between the motion and the electronic state during each gate. Keeping it low is essential not only for immediate gate fidelity but also for preventing motional heating. If each gate adds even a small amount of energy to the motional mode, the system will gradually heat over the course of many gates, limiting circuit depth and fidelity unless periodic recooling is possible.\\
Understanding and mitigating these motion-induced gate errors is central to scalable, high-fidelity atomic quantum computing.
\subsection{Neutral-atom case studies}
Neutral-atom processors use alkali or alkaline-earth atoms trapped in optical tweezers or lattices, and support both analog and digital operation.\\
In the analog mode, the system evolves continuously under the influence of shaped laser fields, enabling the simulation of complex many-body quantum dynamics or the realization of adiabatic quantum optimization. The digital mode, by contrast, is based on discrete quantum gate operations, allowing the construction of universal quantum circuits from sequences of single- and multi-qubit gates.\\
A key feature of neutral atom platforms is the ability to engineer interactions between spin qubits by coupling them to highly excited Rydberg states. Two main types of interaction are utilized. Pair interactions arise when both atoms are simultaneously excited to Rydberg states, resulting in a dipole--dipole coupling that enables resonant excitation exchange, which is central to certain entangling gate schemes. Van der Waals interactions, on the other hand, dominate when only one atom is excited. These interactions, which scale as $1/R^6$ with the interatomic distance $R$, induce energy shifts that underlie the Rydberg blockade effect used in conditional gate protocols.\\
Single-qubit gates are realized using coherent laser pulses that drive transitions between the qubit states, which are often hyperfine ground states. These gates benefit from long coherence times and high-fidelity control enabled by precise optical addressing.\\
Two-qubit gates exploit the Rydberg blockade mechanism: when one atom is excited to a Rydberg state, the resulting interaction shifts the energy levels of nearby atoms, effectively preventing their simultaneous excitation. This blockade enables the implementation of conditional operations such as controlled-Z and controlled-NOT gates by selectively coupling the computational states to the Rydberg level. These gates are fast, inherently entangling, and critical for achieving scalable quantum logic.\\
In summary, neutral atom quantum processors combine tunable interactions, precise optical control, and the ability to operate in both analog and digital regimes, making them a versatile and competitive platform for quantum computing.\\
In the following section, we present applications for a single-qubit gate that is recoil-free and robust against entanglement and heating, as well as a two-qubit gate. Additionally, we include a quantum analog application using real hardware. 

\subsubsection{Recoil-free gates}
\label{sec:recoil_free_gate}
We examine a system consisting of a qubit coupled to a harmonic oscillator, representing the motional state. Our focus is on the physics of \(^{88}\)Sr. The atom is trapped along the direction of the driving laser, with a trapping frequency denoted by $\omega$. \\
We assume perfect ground-state preparation for the qubit, while the motional state is initially in a thermal distribution characterized by a ground-state occupation probability \( p_0 \). Laser-intensity inhomogeneities introduce a relative intensity fluctuation \(\Delta I/I\), which affects both the detuning and the driving terms in the system Hamiltonian~\cite{taichenachev2006magnetic, zhang2024recoil}.
The resulting Hamiltonian describing the dynamics of a trapped atom driven by the clock laser beam is
\begin{align*}
    \frac{H(t)}{\hbar} &= \omega(1 + \delta\omega)\, a^{\dagger}a - \delta(t)\, \sigma_{z} \\
    & \quad + \frac{\Omega(t)}{2} \left( \sigma_{+} e^{i \eta(a^{\dagger}+a) + i\phi(t)} + \text{h.c.} \right)
\end{align*}
Here $\omega$ is the motional trap frequency, $\delta\omega$ its relative fluctuation, and $\delta(t)\sigma_z$ is the detuning contribution to the qubit Hamiltonian. The final term drives the transition; $\eta$ is the Lamb-Dicke parameter quantifying photon-recoil coupling between internal and motional states.\\
The Rabi frequency, $\Omega(t)$, which drives the transition, is proportional to the square root of the laser intensity, $I(t)$, and the magnetic field at the atom's location, $B(t)$, given by $\Omega(t) = \kappa \sqrt{I(t)} B(t)$.\\
The Rabi frequency is expressed as 
\begin{equation}
\Omega(t) = \Omega_{\text{max}} f(t) \left( 1 + \frac{1}{2} \frac{\Delta I}{I} \right),
\end{equation}
where $\Omega_{\text{max}}$ is the maximum Rabi frequency, $f(t)$ is a dimensionless pulse shape bounded between 0 and 1, and $\Delta I/I$ is the relative intensity fluctuation. \\
The detuning, $\delta(t)$, includes the AC Stark shift arising from intensity fluctuations and is modeled as
\begin{equation}
\delta(t) = \frac{\alpha}{2 c \varepsilon_0 K^2} \frac{\Omega^{2}(t)}{B^2}\,  \frac{\Delta I}{I}.
\end{equation}
Here, $\alpha$ is the differential polarizability between the ${}^1S_0$ ground state and ${}^3P_0$ clock state at the clock transition wavelength, $B$ is the magnetic-field amplitude, $K$ is the proportionality constant relating the Rabi frequency to the applied field, $c$ is the speed of light, and $\varepsilon_0$ is the vacuum permittivity.\\
The phase $\phi(t)$ appears explicitly in the drive factor $e^{i\phi(t)}$ in the Hamiltonian and is the optimized control field in this protocol. 
\paragraph{Quantum process fidelity}
For a composite qubit--motion system, the full evolution operator does not in general define a unitary operation on the qubit alone: residual qubit--motion entanglement makes the reduced qubit dynamics non-unitary. We therefore use quantum process tomography (QPT) to quantify the qubit operation and residual entanglement with motion.\\
To evaluate the single-qubit gate performance, we define the QPT-based fidelity-error cost function (Appendix \ref{app:recoil-free-protocol}):
\begin{equation}
1 - \langle \mathcal{F} \rangle \approx 
\underbrace{\frac{2}{3}(1 - \chi_0)}_{J_{\text{entanglement}}} + 
\underbrace{\frac{2}{3}\left( 1 - \left| \operatorname{Tr} \left( U_{\text{target}}^{\dagger} \hat{E}_0 \right) / n \right|^2 \right)}_{J_{\text{unitarity}}}
\end{equation}
Here $\langle\mathcal{F}\rangle$ is the average gate fidelity, $\chi_0$ is the dominant process-matrix eigenvalue, $\hat{E}_0$ is its associated effective qubit operator, $n=2$ is the qubit Hilbert-space dimension, and $U_{\text{target}}$ is the target unitary. The first term, \( J_{\text{entanglement}} \), captures non-unitarity induced by residual qubit--motion entanglement; the second, \( J_{\text{unitarity}} \), measures deviation of the projected qubit operation from $U_{\text{target}}$.

\begin{figure}
    \centering
    \includegraphics[width=0.9\linewidth]{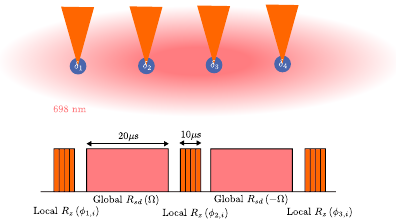}
    \caption{The recoil-free protocol consists of two global optimized pulses (each of duration 20~$\mu$s) and local $Z$-rotations (each of approximately 1~$\mu$s).}
    \label{fig:recoil-free-protocol}
\end{figure}

\paragraph{Recoil-free protocol}
To realize an arbitrary single-qubit rotation, \( R_x(\theta) \), we decompose it into a sequence of $X$ and $Z$ rotations:
\begin{equation*}
    R_x(\theta) = R_z(\theta_1) R_x(\pi/2) R_z(\theta_2) R_x(-\pi/2) R_z(\theta_3)
\end{equation*}
The recoil-free protocol implements this decomposition with two optimized global $X$-rotation pulses and three local $Z$-rotations, as shown in \autoref{fig:recoil-free-protocol}. The global pulses were optimized to suppress motional entanglement and recoil-induced heating while implementing the required equatorial rotations. Robust QOC was used to make the pulses resilient to detuning inhomogeneities.\\
The local $Z$-rotations are then used to reach arbitrary points on the Bloch sphere. These are implemented using fast, high-fidelity techniques, possibly leveraging local light shifts or microwave fields.\\
The recoil-free protocol executes thousands of high-fidelity single-qubit gates, with overall fidelities exceeding 99\%~\cite{zhang2024recoil}.
\subsubsection{Controlled-Z gate}
\begin{figure}[ht!]
    \centering
    \includegraphics[trim={1.5cm 0.5cm 0.5cm 0.5cm},clip, width=0.9\linewidth]{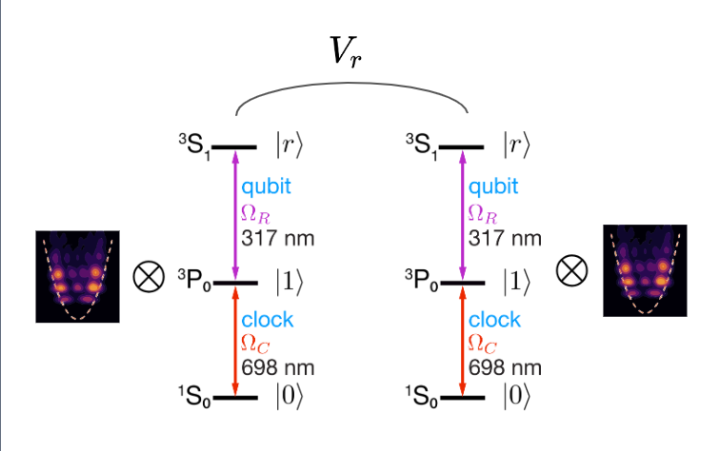}
    \caption{Level diagram for the two-qubit Rydberg gate. The qubit states are coupled to the Rydberg level via laser pulses with time-dependent Rabi frequency and phase.}
    \label{fig:2-qubit-schema}
\end{figure}

\begin{figure}[ht!]
    \centering
    \includegraphics[width=0.98\linewidth]{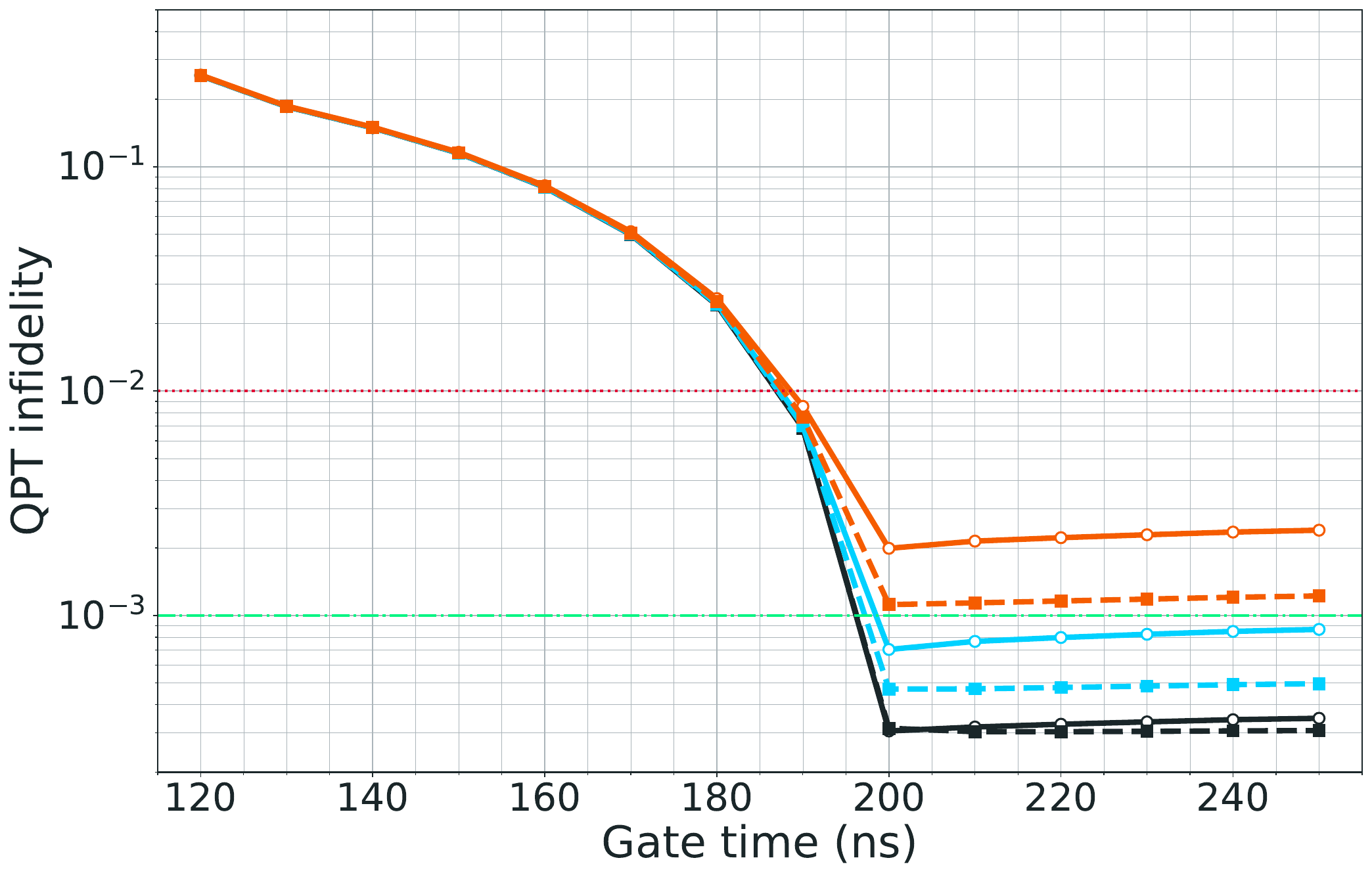}
    \caption{QPT infidelity versus gate duration for the CZ-gate sweep. Each curve represents one combination of initial motional ground-state occupation $p_0$ and trap frequency $\omega_{\mathrm{trap}}$; color denotes $p_0 \in [0.99,\,0.95,\,0.90]$ respectively black, green, and orange, while line style and marker distinguish trap frequencies $[75,\,125]\text{ KHz}$ respectively solid and dashed. The red and green horizontal lines mark infidelity thresholds of 1\% and 0.1\%, respectively.}
    \label{fig:quantum-speed-limit}
\end{figure}

\begin{figure}[ht!]
    \centering
    \includegraphics[width=0.98\linewidth]{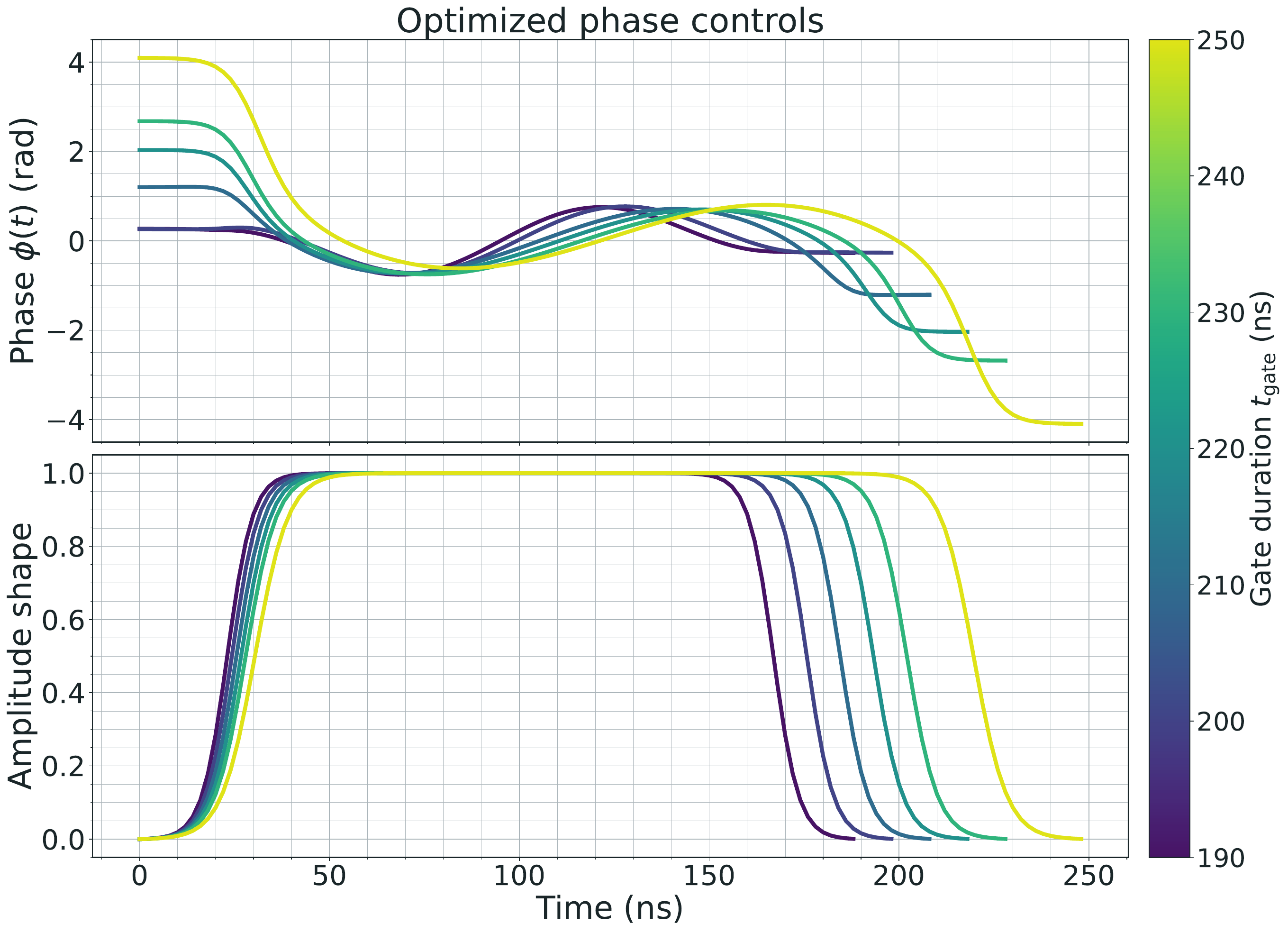}
    \caption{Optimized control signals for CZ gates with durations above the sampled quantum speed limit. The upper panel shows the phase control $\phi(t)$ in radians; the lower panel shows the Rabi-frequency envelope $\Omega(t)/2\pi$ in MHz. Each colored curve corresponds to the gate duration indicated in the legend.}
    \label{fig:optimized_pulses-2-qubit}
\end{figure}

We consider two qubits trapped at a triple-magic wavelength, for which the relevant clock and Rydberg states experience matched trapping conditions, with trap frequency \(\omega\) along the driving-laser direction.
The interaction potential between the Rydberg states is governed by the Van der Waals interaction, characterized by a strength \(V_r\) at an interatomic distance \(d_{\text{atom}}\).
Each qubit is coupled to its respective motional degrees of freedom (\autoref{fig:2-qubit-schema}).\\
The system is initially prepared as a tensor product of the qubit ground states, while the motional state follows a thermal distribution according to the Boltzmann distribution, with a ground-state occupation probability \( p_0 \).\\
The Hamiltonian for the two-qubit gate is
\begin{equation*}
\begin{aligned}
H_r = &\sum_{i=1}^2 \Bigg[ \Delta \ket{r_i}\bra{r_i} \\
&+ \frac{\Omega_{\max}}{2} f(t) \Big( \ket{1_i}\bra{r_i} \, e^{i \phi(t) + i \sum_{\alpha} \eta_\alpha (a_{i,\alpha}^\dagger + a_{i,\alpha})} \\
& \quad + \text{h.c.} \Big) \Bigg] + V \ket{r_1 r_2}\bra{r_2 r_1}
\end{aligned}
\end{equation*}
where $\ket{r_i}$ is the Rydberg state of atom $i$, $a_{i,\alpha}$ is the motional annihilation operator along $\alpha\in\{x,y,z\}$, $\eta_\alpha$ is its Lamb-Dicke parameter, and $V$ is the Rydberg--Rydberg interaction strength.\\
Two atoms prepared in Rydberg states will experience a repulsive force due to the Van der Waals potential.
The Hamiltonian under linear approximation around the equilibrium position is
\begin{equation*}
\begin{aligned}
        H_{rf}& =-\frac{6 V}{d_{\text{atom }}}\ket{r_1 r_2} \bra{r_2 r_1} \\ & \sum_{\alpha} \sqrt{\frac{\hbar}{2 M \omega_\alpha}}\left(a_{2, \alpha}^{\dagger}+a_{2, \alpha}-a_{1, \alpha}^{\dagger}-a_{1, \alpha}\right)
\end{aligned}
\end{equation*}
To take into account the decay from the Rydberg states, we introduce a non-Hermitian term contribution to the Hamiltonian with
\begin{equation}
    H_{\text{decay}}=-\frac{i}{2} \Gamma_r \sum_{i=1,2}\ket{r_i} \bra{r_i}
\end{equation}
where $M$ is the atomic mass and $\Gamma_r=1/(100~\mu\mathrm{s})$ is the Rydberg-state decay rate.\\
\paragraph{Geometric quantum gate}
We implement a geometric CZ gate by shaping the phase control, $\phi(t)$, through optimal control techniques. The optimization targets the total infidelity, incorporating the dominant noise sources of the system (see Appendix~\ref{app:infidelities-2-qubit-gate}). A smooth envelope is used for the Rabi frequency, inspired by recent advances in high-fidelity gate design~\cite{levine2019parallel, jandura2022time, pagano2022error}.\\
\autoref{fig:quantum-speed-limit} shows the simulated infidelity as a function of scaled gate duration for the modeled motional occupations and trap frequencies. The curves approach a performance plateau beyond $\Omega_{\max}T\approx1.61$, identifying the operating range used for the optimized phase and amplitude controls in \autoref{fig:optimized_pulses-2-qubit}. Quantitative comparison to experimental results remains pending replacement of the source figure.

\subsubsection{Quantum simulation with Aquila}
We used Aquila, QuEra's programmable neutral-atom quantum processor, for analog quantum simulation. AWS Braket provided the interface to its analog classical simulator and to the Aquila QPU.\\
We aim to generate an antiferromagnetic state within a square lattice of Rydberg atom arrays. The Hamiltonian we aim to simulate is expressed as follows:

$$
H=\frac{\Omega(t)}{2} \sum_i \sigma_x^{(i)}-\Delta(t) \sum_i n_i+\sum_{i<j} \frac{V}{\left|r_i-r_j\right|^6} n_i n_j
$$

\begin{figure}[!htb]
    \centering
    \begin{subfigure}[h]{0.7\linewidth}
        \centering
        \includegraphics[width=\linewidth]{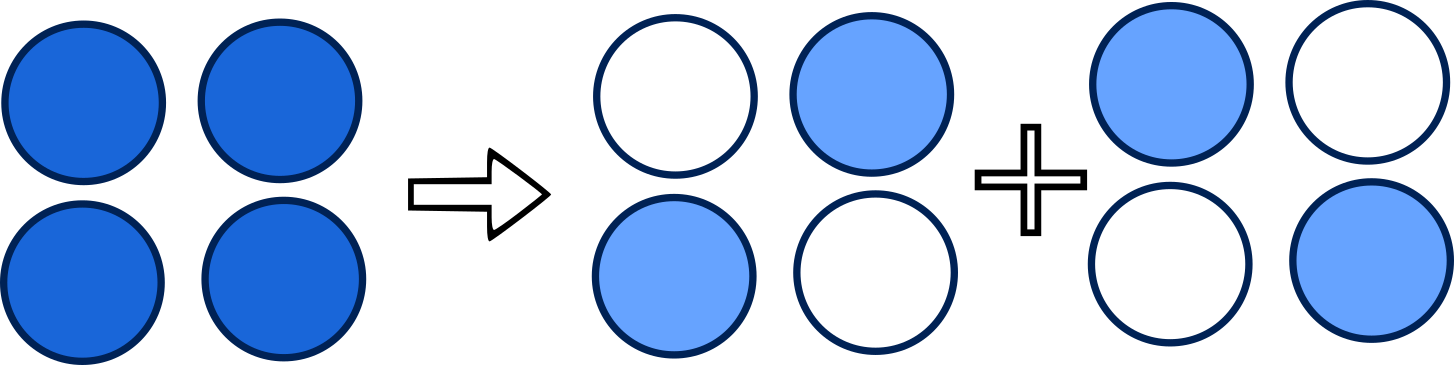}
        \caption{Schematic antiferromagnetic target state, with alternating ground and Rydberg occupations.}
        \label{fig:af-state}
    \end{subfigure}
    \begin{subfigure}[h]{0.8\linewidth}
        \centering
        \includegraphics[width=\linewidth]{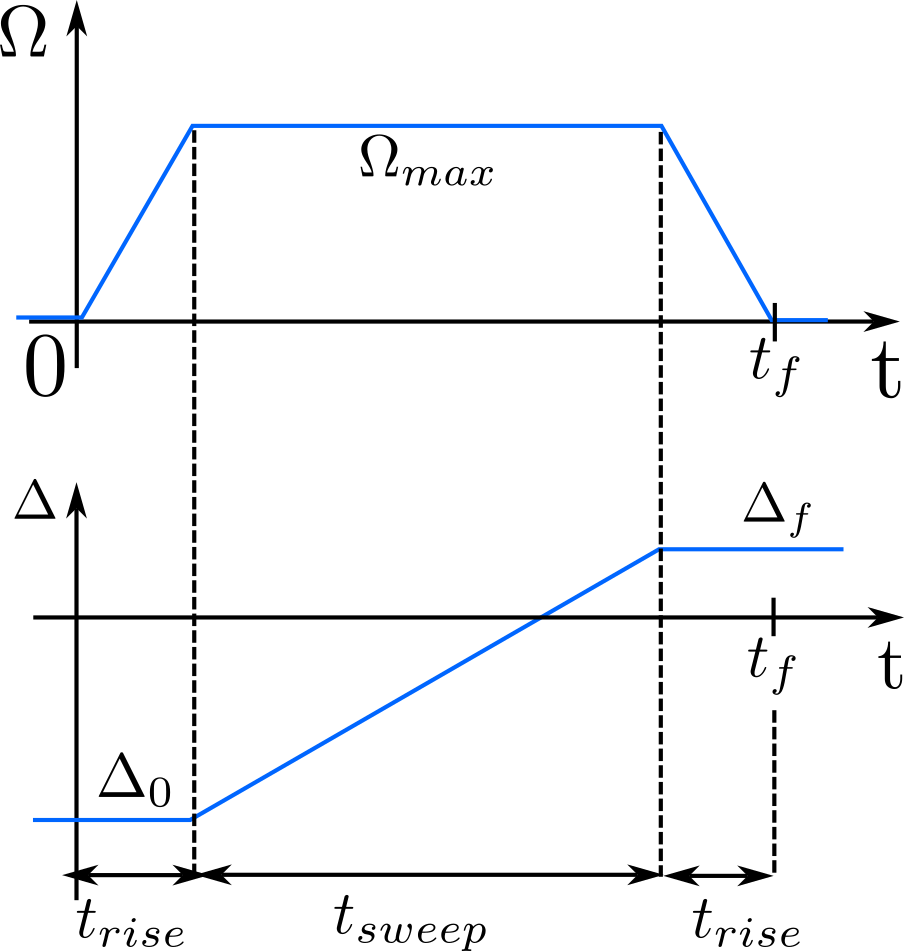}
        \caption{Adiabatic control sweep of Rabi amplitude and detuning used for baseline state preparation.}
        \label{fig:adiabatic-sweep}
    \end{subfigure}
\end{figure}

In this Hamiltonian, the spin operators act on the $i^\text{th}$ spin, and the terms represent the Rabi amplitude, the detuning controls, and the Van der Waals interaction, respectively.\\
The target is an antiferromagnetic state with alternating ground and Rydberg occupations, illustrated schematically in \autoref{fig:af-state}.
\begin{figure*}[!t]
    \centering
    \includegraphics[width=0.6\textwidth]{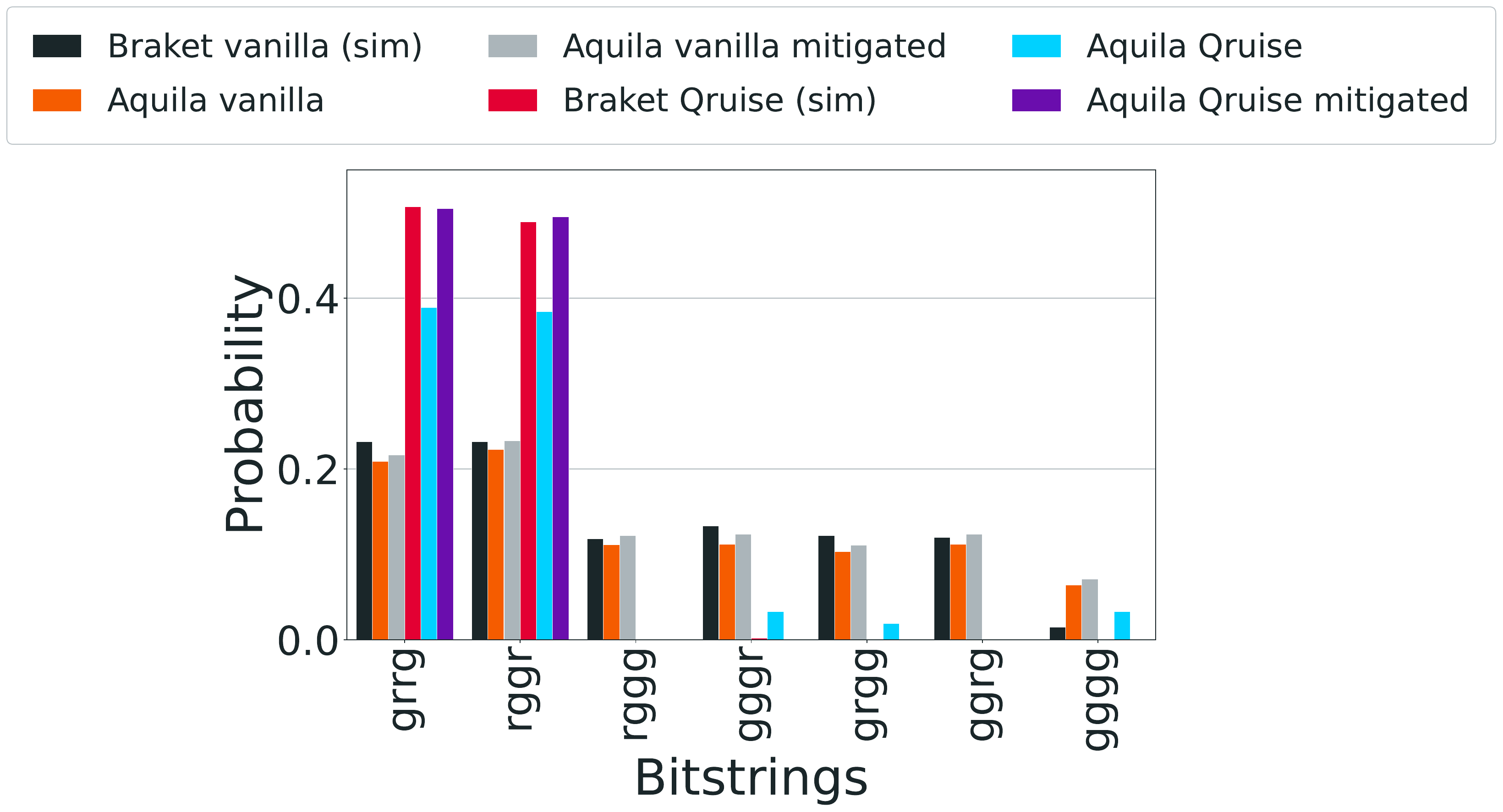}
    \caption{Output probabilities for the seven four-atom bitstrings shown on the horizontal axis; $g$ and $r$ denote ground and Rydberg states. Within each bitstring group, bars from left to right represent: Braket vanilla simulation, unmitigated Aquila vanilla, readout-mitigated Aquila vanilla, Braket simulation with Qruise controls, unmitigated Aquila with Qruise controls, and readout-mitigated Aquila with Qruise controls. The vertical axis shows the probability assigned to each bitstring.}
    \label{fig:aquila_results}
\end{figure*}

Baseline preparation used an adiabatic sweep of the detuning and Rabi amplitude, shown in \autoref{fig:adiabatic-sweep}. In \autoref{fig:aquila_results}, the Braket baseline histograms were produced by the AWS Braket simulator, while the Aquila vanilla baseline was measured on the QPU.
The figure shows that the state produced by this protocol is not fully antiferromagnetic, as significant probabilities associated with other spin states remain.
\paragraph{Optimization and error mitigation}
To optimize the generation of the antiferromagnetic state, we employed optimal control techniques using the  \texttt{qruise-toolset} software to fine-tune the Rabi amplitude and detuning. The optimized controls were tested using AWS Braket's analog classical simulator and subsequently implemented on the Aquila QPU. During optimization, the controls were constrained by the QPU bandwidth and time resolution. Post-processing used readout-error information to mitigate the measured histograms, showing a fully antiferromagnetic state as shown in \autoref{fig:aquila_results} for Braket Qruise simulation and the Aquila Qruise mitigated bars.

\subsection{Trapped ions}
Trapped-ion quantum computing stands as one of the most promising platforms for quantum computation, renowned for its long coherence times, high gate fidelities, and precise qubit control. In these systems, qubits are encoded in the computational subspace of the electronic ground state and a metastable excited state of ions, typically alkaline-earth or alkali-like atoms. These ions are confined in a radio frequency (RF) surface trap, forming a Coulomb crystal whose collective dynamics is governed by harmonic oscillator modes. The center-of-mass mode, characterized by the lowest frequency, plays a central role in quantum gate operations due to its strong and uniform coupling across the ion chain.\\
Among the earliest quantum-computing proposals was the Cirac--Zoller scheme based on individually trapped ions \cite{Cirac1995PRL}. In RF traps, collective motional modes provide a quantum bus that mediates entanglement between ion qubits \cite{Bruzewicz2019TrappedionQC}. \\
Ion qubits offer intrinsic advantages for quantum computing implementations: they are fundamentally identical (like neutral-atom qubits), exhibit very long coherence times (up to hours), and enable near-perfect qubit-state initialization and detection.
In 2000, DiVincenzo \cite{DiVincenzo2000} published five key criteria for a quantum information processor. Trapped ions represent one of only a few qubit technologies which have yet to fulfil all of DiVincenzo's original criteria with high fidelity.\\
\paragraph{System characteristics}
Ions are trapped in an electromagnetic potential under ultra-high vacuum, where the ions repel each other via Coulomb interaction under harmonic electromagnetic force, and are well isolated from their environment.
Ions are generally trapped with a DC static potential and an oscillating RF pseudo-potential.
These ions are generally cooled via Doppler cooling and loaded into the trap under the RF potential. Since the RF pseudo-potential is approximately one order of magnitude higher than the static DC potential, the ions form a linear chain along the $z$-direction. \\
Several types of qubits can be implemented in ion-based platforms, depending on the specific internal states chosen to define the qubit. Each type exhibits distinct characteristics: Zeeman qubits (with splittings of tens of MHz), hyperfine qubits (tens of GHz), and optical qubits (hundreds of THz)

\subsubsection{Example: M\o lmer-S\o rensen gate simulation and optimal control}
\begin{figure}[!htb]
    \centering
    \includegraphics[width=0.6\linewidth]{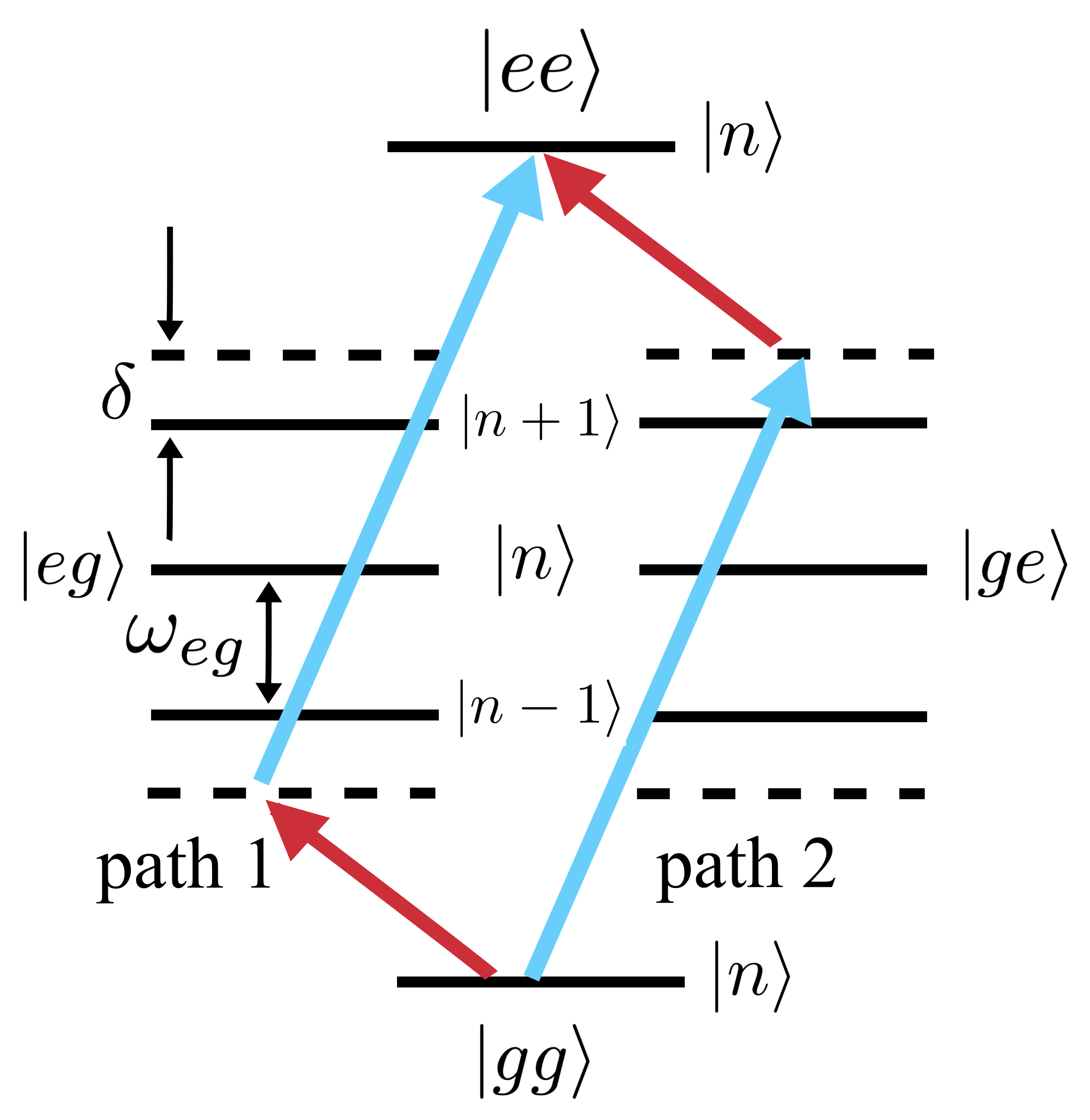}
    \caption{Representative phase-space trajectories for an MS gate \cite{Cai2023}. A closed trajectory returns the shared motional mode to its initial state while the enclosed area produces the entangling geometric phase.}
    \label{fig:MSgateScheme}
\end{figure}

The Mølmer--Sørensen (MS) gate is a trapped-ion entangling gate based on a shared vibrational mode. To first order, it is insensitive to the initial temperature of that mode and can entangle more than two ions in a single operation. \autoref{fig:MSgateScheme} illustrates the closed phase-space trajectories that disentangle the motion at the end of an ideal gate.

To implement the gate, two ions are irradiated with a bichromatic laser field with frequencies $\omega_{e g} \pm \delta$, where $\hbar \omega_{e g}$ is the energy splitting of the qubit states, and $\delta$ is a detuning close to the motional frequency of the ions. Depending on the interaction time, this produces the states:
$$
\begin{aligned}
|e e\rangle & \rightarrow(|e e\rangle+i|g g\rangle) / \sqrt{2} \\
|e g\rangle & \rightarrow(|e g\rangle-i|g e\rangle) / \sqrt{2} \\
|g e\rangle & \rightarrow(|g e\rangle-i|e g\rangle) / \sqrt{2} \\
|g g\rangle & \rightarrow(|g g\rangle+i|e e\rangle) / \sqrt{2}
\end{aligned}
$$
The above is equivalent to the Ising coupling gate $R_{y y}(\pi / 2)$. It can be shown that this gate (along with arbitrary single-qubit rotation) produces a universal set of gates.\\
For the MS gate simulation definition, we use two ions and a common motional mode. For the drive, we'll use a general interaction Hamiltonian for one qubit and one motional state (\cite{Cai2023} Eq.3) that reads as:

\begin{align*}
    \hat{H}_{int}=\frac{\hbar}{2} \Omega \hat{\sigma}^{+} \left[ 1 + i \eta ( \hat{a}^{\dagger} e^{ i \omega_m t} + \hat{a} e^{ -i \omega_m t} ) \right] e^{i (\phi - \delta t)} \\
    + \text{h.c.}
\end{align*}

The Hamiltonian of the system in the interaction picture and with the bichromatic light, after applying the rotating wave approximation, is given by:

\begin{equation}
\hat{H}_{MS}= \sum^{2}_{j=1} \eta \frac{\hbar \Omega}{2} \left(\hat{a}^{\dag} e^{- 
\xi_0 t} e^{i \phi} + \hat{a} e^{ 
\xi_0 t} e^{-i \phi}\right) \hat{\sigma}^{\phi_S}_j 
\label{MSHamiltonian}
\end{equation}

\begin{figure}[!htb]
    \centering
    \begin{subfigure}[t]{\linewidth}
    \centering
    \includegraphics[width=0.82\linewidth]{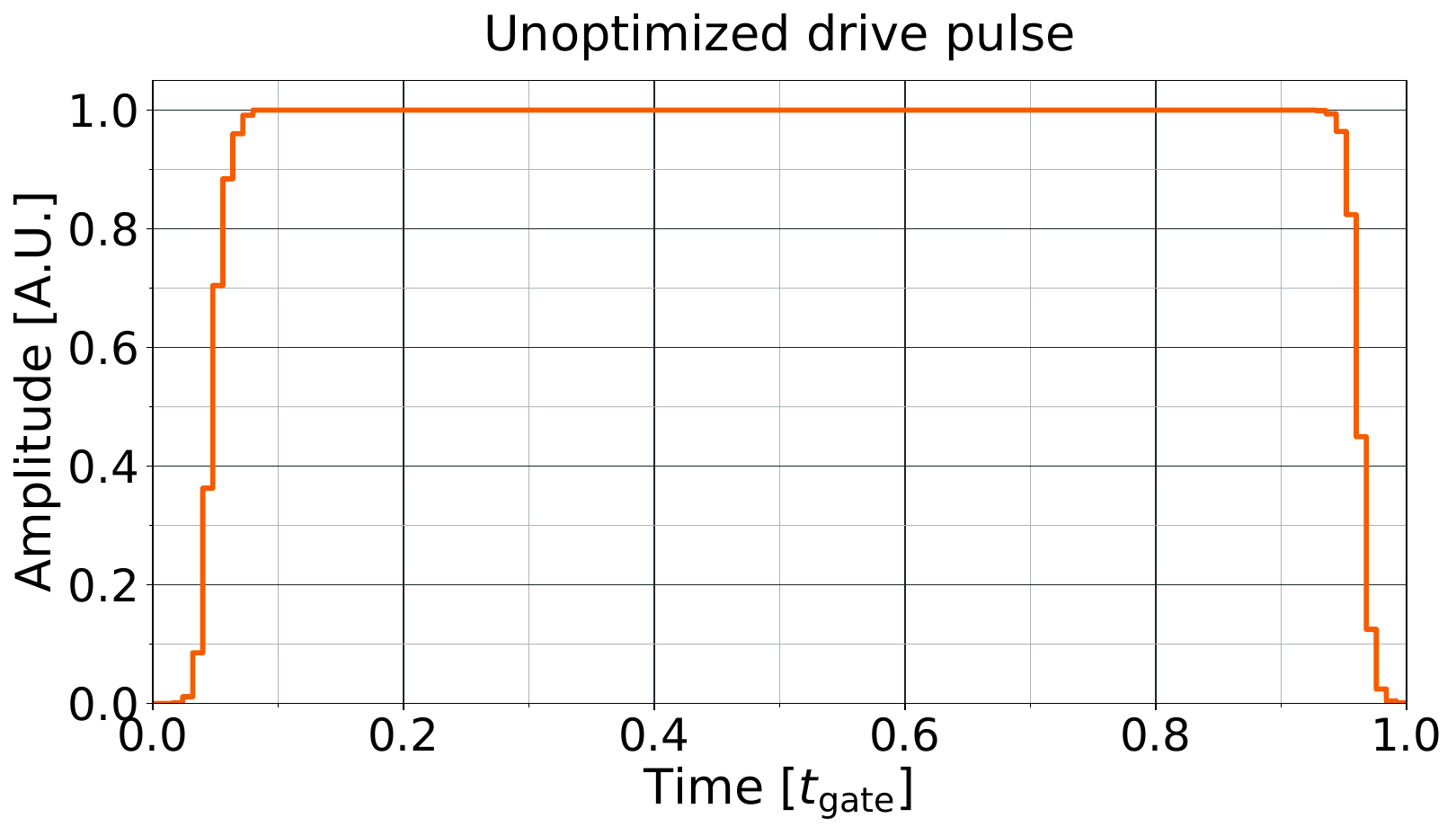}
    \caption{Normalized square-amplitude drive pulse.}
    \end{subfigure}\par\medskip
    \begin{subfigure}[t]{\linewidth}
    \centering
    \includegraphics[width=0.82\linewidth]{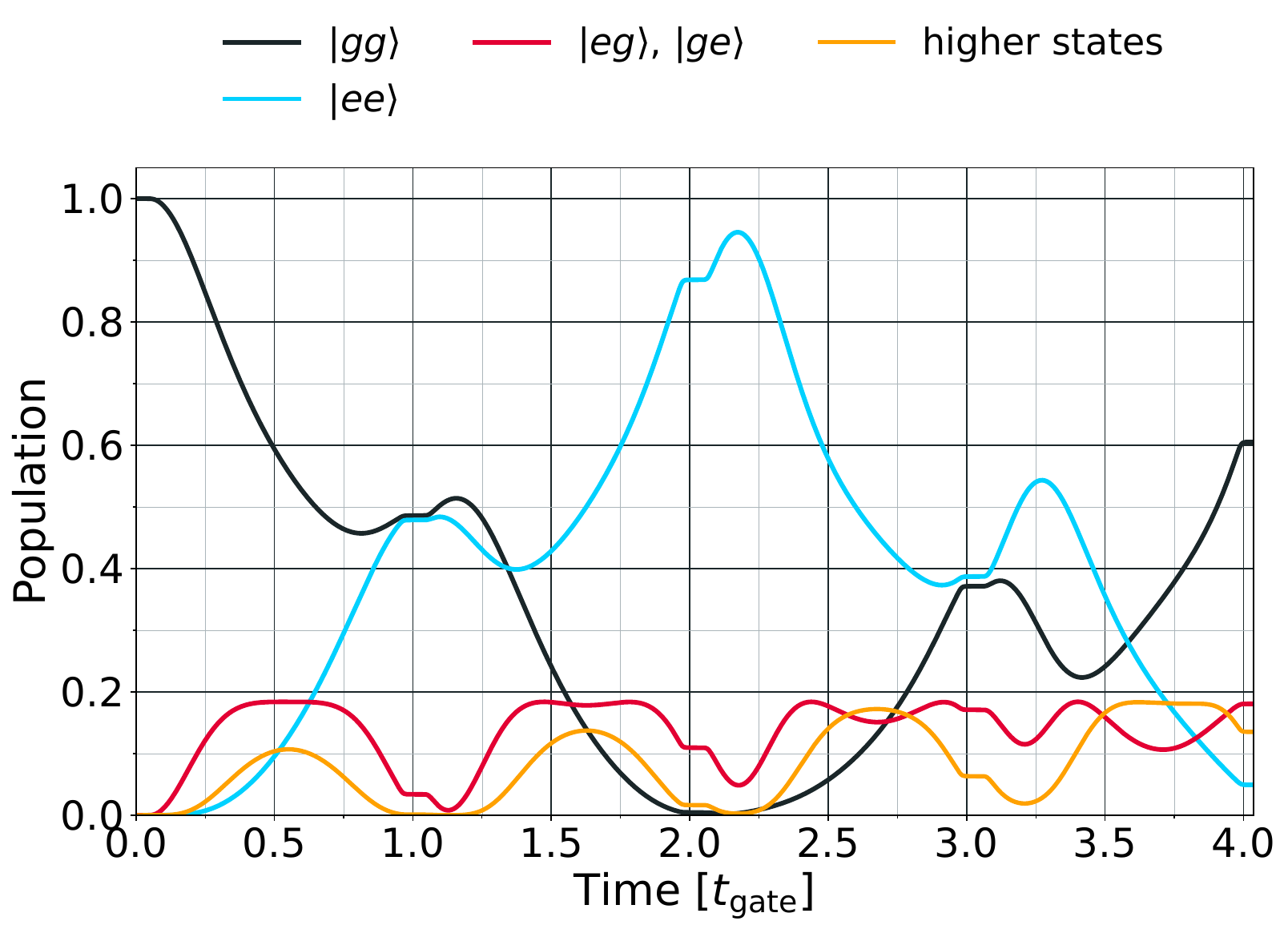}
    \caption{State populations over four consecutive gate durations.}
    \end{subfigure}
    \caption{Unoptimized Mølmer--Sørensen gate. The upper panel shows the normalized Rabi-amplitude control sampled at the 10~MHz AWG resolution. The lower panel shows the populations of the computational states and the combined single-phonon states; the remaining motional-state populations are grouped as higher states. The simulated gate infidelity is $0.036$.}
    \label{fig:MSNonOptimizedSignalandPopulation}
\end{figure}

Here $\omega_m$ is the shared motional-mode frequency, $\eta$ the Lamb-Dicke parameter, $\Omega$ the Rabi frequency, $\hat a$ and $\hat a^\dagger$ the phonon annihilation and creation operators, and $\hat\sigma_j^{\phi_S}$ the spin operator for ion $j$. The bichromatic tones are detuned from the blue and red sidebands by $+\xi_0$ and $-\xi_0$, respectively; the signs follow from the sideband frequencies $\omega_{eg}\pm\omega_m$. The phase-dependent spin operator is $\hat{\sigma}^{\phi_S} = \hat{\sigma}^x
\cos{\phi_S} + \hat{\sigma}^y \sin{\phi_S}$, where $\phi_S$ correlates with the qubit state.\\
The gate duration $t_{\mathrm{gate}}$ is selected so that the motional trajectory closes and the accumulated entangling phase reaches its target value (Appendix \ref{app:ms-gate-optimal-parameters}). \autoref{fig:MSNonOptimizedSignalandPopulation} shows the square-amplitude drive and the resulting state populations over four consecutive gates, with total duration $4t_{\mathrm{gate}}$. The drive envelope is plotted at the 10~MHz AWG sampling resolution; the population trajectories are sampled from the saved numerical simulation output.

\begin{figure}[!htb]
\centering
\begin{subfigure}[t]{\linewidth}
\centering
\includegraphics[width=0.82\linewidth]{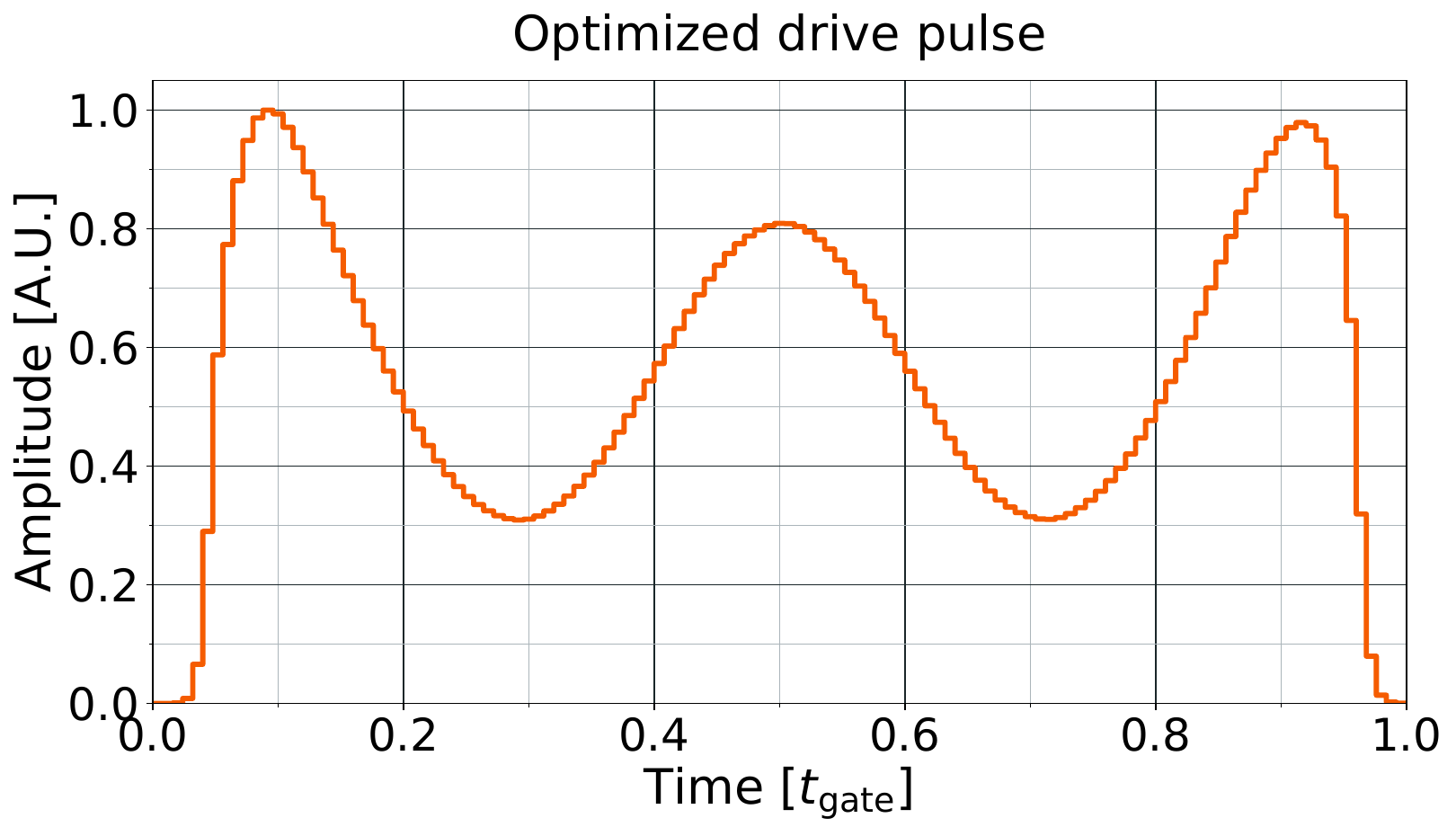}
\caption{Normalized optimized drive pulse.}
\end{subfigure}\par\medskip
\begin{subfigure}[t]{\linewidth}
\centering
\includegraphics[width=0.82\linewidth]{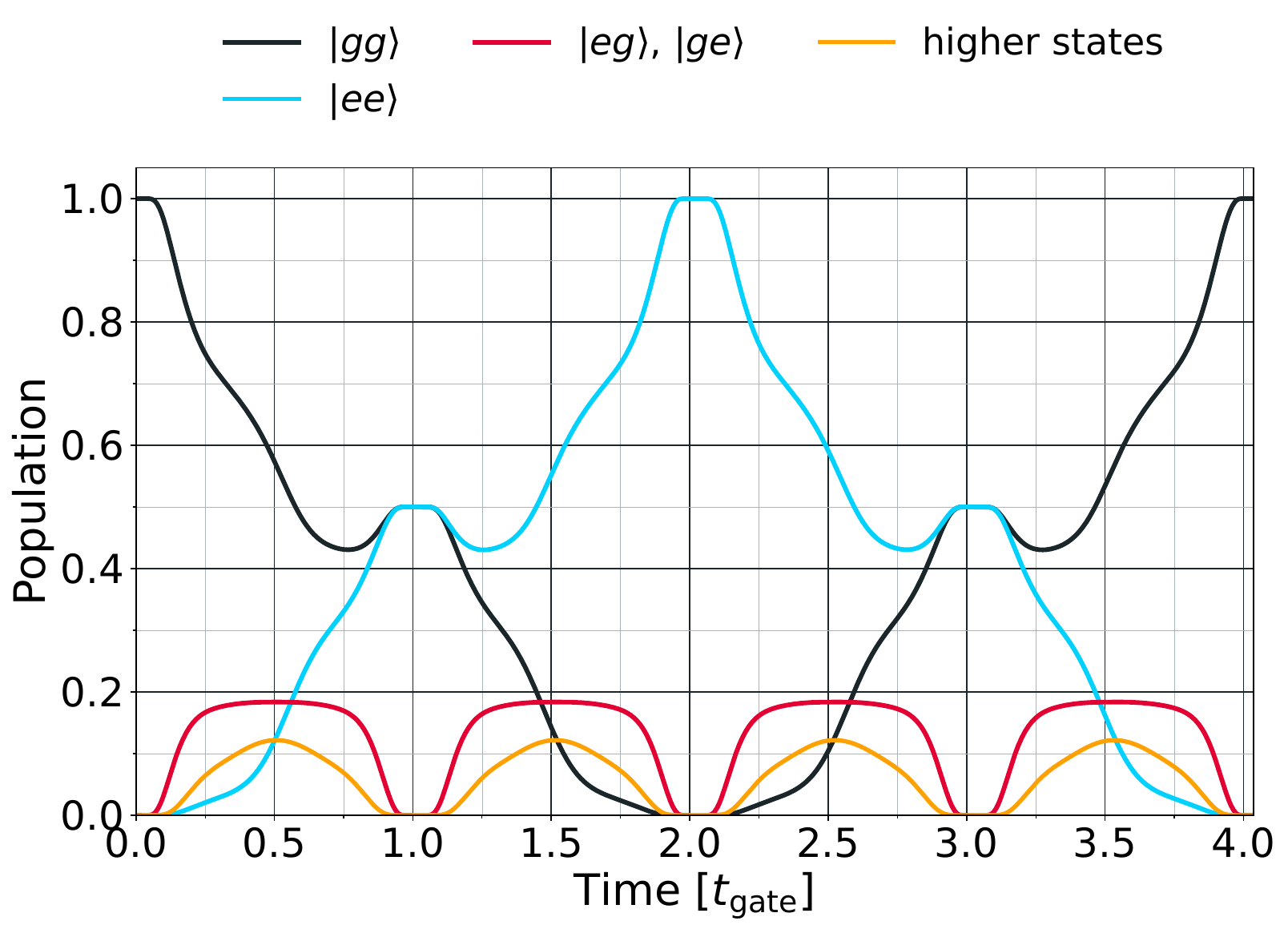}
\caption{State populations over four consecutive gate durations.}
\end{subfigure}
\caption{Optimized Mølmer--Sørensen gate. The upper panel shows the optimized normalized Rabi-amplitude control at 10~MHz AWG resolution. The lower panel shows the populations of the computational states and combined single-phonon states; the remaining motional-state populations are grouped as higher states. The saved simulation reports a gate infidelity of $1.59\times10^{-5}$.}

\label{fig:MSOptimizedSignalandPopulation}
\end{figure}

\paragraph{Optimization}
Control-electronics imperfections reduce the gate fidelity. QOC therefore optimized the time-dependent Rabi amplitude $\Omega(t)$ to compensate for the modeled distortions.
The final, optimized pulse shape is shown in \autoref{fig:MSOptimizedSignalandPopulation}, together with the population dynamics for that pulse over $4t_{\mathrm{gate}}$. The optimized control compensates for the modeled control-electronics response and produces a gate infidelity of $1.59\times10^{-5}$ in the saved simulation. The pulse is shown at the 10~MHz AWG sampling resolution.
\subsubsection{Light-shift gate}

A two-qubit entangling gate scheme was proposed by Milburn in 2000 \cite{Milburn2000} and demonstrated with $^{9}\text{Be}^{+}$ hyperfine qubits, which is the so-called light-shift (LS) gate. Two qubits in a common bus mode where both ions are illuminated simultaneously.
Similar to the Mølmer–Sørensen gate scheme, the LS gate is driven by a state-dependent force, and generates the entangling interaction between ion qubits with the collective vibrational mode. However, the $\ket{0} \leftrightarrow \ket{1}$ transition is not required during the LS gate. The LS gate is achieved by applying two beams with a frequency difference close to a target vibrational mode, with the coupling acting along the direction of the wavevector difference.
This geometric phase gate can be executed fast as the laser can be tuned such that off-resonant transitions are unlikely. The Cirac-Zoller and M\o lmer-S\o rensen gates, on the other hand, require a laser tuned to or close to a sideband transition. \\

\begin{figure}
    \centering
    \includegraphics[width=0.6\linewidth]{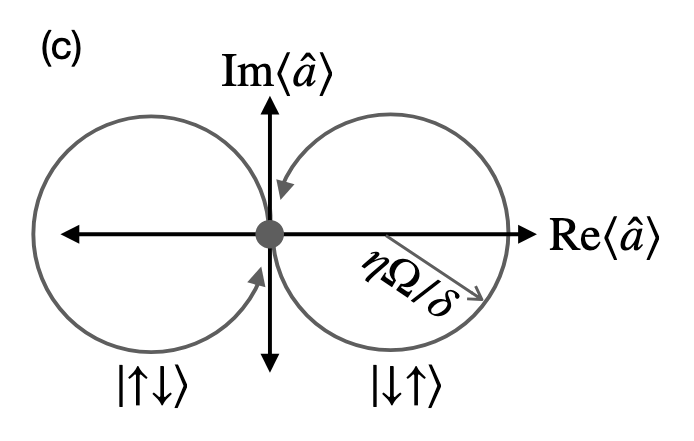}
    \caption{ $\ket{01}$
and $\ket{10}$ states are driven, undergoing circular trajectories in phase space that return to their origin at a time $t_{gate}$.}
    \label{fig:phase_plot_honeywell}
\end{figure}

In the LS gate, two non-co-propagating laser fields (red and blue-detuned) create a standing wave.
Equal and opposite detunings cancel out average AC-Stark shifts that would decohere the system.
The ion is driven back to the original motional state after the full gate time. This ensures that the motion is disentangled from the electronic state after the gate operation. This is shown schematically in
\autoref{fig:phase_plot_honeywell}.

In the Lamb-Dicke regime and the rotating wave approximation, the interaction Hamiltonian can be written as:

\begin{equation*}
\hat{H}_{LS}= \sum^{2}_{j=1} \eta \frac{\hbar \Omega}{2} \left(\hat{a}^{\dagger} e^{- 
\xi_0 t} e^{i \phi} + \hat{a} e^{ 
\xi_0 t} e^{-i \phi}\right) \hat{\sigma}^{z}_j 
\end{equation*}
where $a^{\dagger}$ and $a$ are the creation and annihilation operators of phonons in the ions' collective motional mode, $\hbar \nu$ is the phonon energy, $\xi_0$ is the driving frequency detuning, and $\hat{\sigma}^{z}_j$ is the Pauli-Z matrix on the $j^\text{th}$ qubit \cite{Cai2023}.\\

\begin{figure}[!htb]
    \centering
    \begin{subfigure}[t]{0.7\linewidth}
        \centering
        \includegraphics[width=\linewidth]{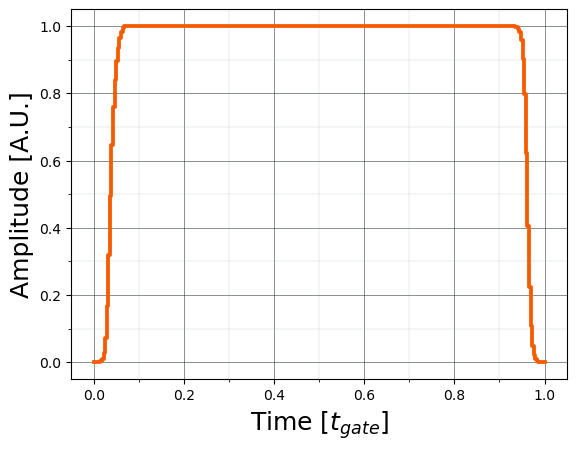}
        \caption{LS gate intensity shape.}
    \end{subfigure}
    \vfill
    \vspace{0.5cm}
    \begin{subfigure}[t]{0.7\linewidth}
        \centering
        \includegraphics[width=\linewidth]{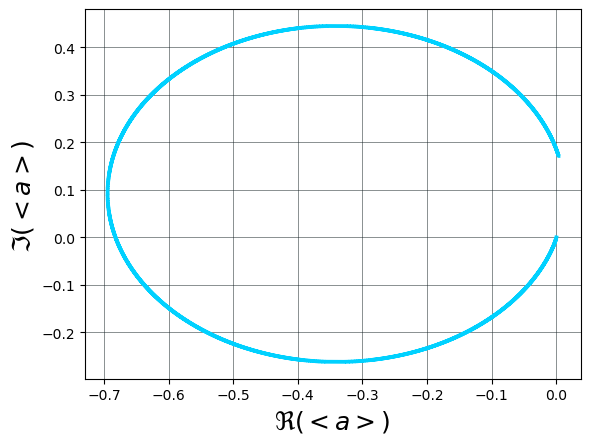}
        \caption{LS gate phase plot.}
    \end{subfigure}
    \caption{LS gate signal and phase dynamics for four times $t_{gate}$. Gate infidelity: $0.0037$ }
    \label{fig:LSNonOptimizedSignalandPopulation}
\end{figure}

After defining the model, the system components, and the drive, we can simulate the system evolution with a square intensity profile for the driving pulse, which is shown in 
\autoref{fig:LSNonOptimizedSignalandPopulation}(a). Subplot (b) shows the phase-space trajectory of the motional state during the gate. We observe that with these conditions, the phase does not return to the same value at the end of the gate, indicating incomplete disentanglement between the motional and electronic states. \\

\begin{figure}[!htb]
    \centering
    \begin{subfigure}[t]{0.7\linewidth}
        \centering
        \includegraphics[width=\linewidth]{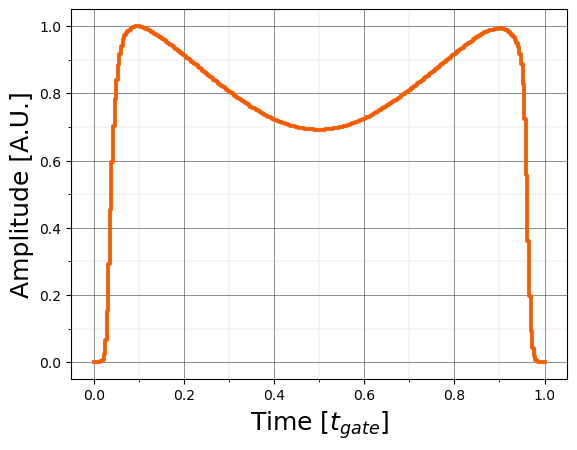}
        \caption{Optimized LS gate intensity shape.}
    \end{subfigure}
    \vfill
    \vspace{0.5cm}
    \begin{subfigure}[t]{0.7\linewidth}
        \centering
        \includegraphics[width=\linewidth]{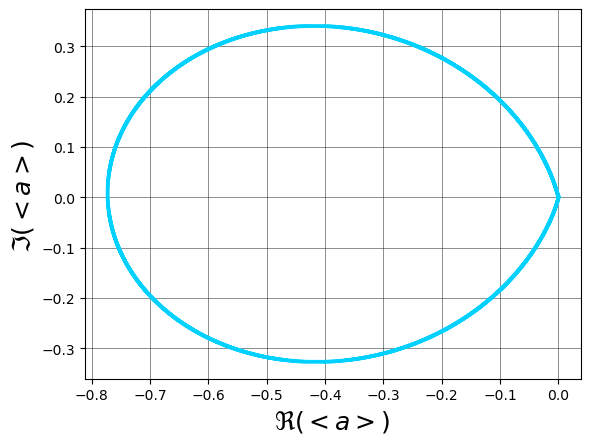}
        \caption{Optimized LS gate phase plot.}
    \end{subfigure}
    \caption{LS gate signal and phase dynamics for four times $t_{gate}$. Gate infidelity: $4.22
e^{-7}$}
    \label{fig:LSOptimizedSignalandPopulation}
\end{figure}

For the optimal control, similar to the MS gate example, we use the AC Stark shift, that depends on beam intensity and detuning, to shape the pulse and find the optimal conditions. 
The improved pulse shape and the new phase trajectory plot can be seen in \autoref{fig:LSOptimizedSignalandPopulation}(a), and (b). In this case, phase trajectory is closed, confirming that the spin and motion are disentangled and the gate is clean.
The fidelity achieved after the optimization greater than $99.99\%$. \\
As a conclusion, we can simulate initial experimental conditions and improve the gate fidelidy with QOC, which brought the gate to have a closed loop in the phase plot, leading to an ideal gate situation after optimization.
\subsection{Bayesian calibration of trapped-ion gates}
\label{subsec:BayesianCalibration}
This section investigates the experimental application of Bayesian techniques for calibrating trapped-ion systems, with a focus on their benefits, challenges, and practical implementation for optimizing gate performance~\cite{PRXQuantum.3.020350}. For theoretical details, see \autoref{subsubsec:BayesianCalibrationTheory}.\\
We consider a system of \( N \) ions confined in a linear Paul trap and driven by a bichromatic laser field composed of two tones with equal phase. Each ion experiences a uniform, time-dependent Rabi frequency, \( \Omega(t) \), allowing for shaped pulses. The transition frequency between the two internal qubit states includes a bare component and an AC Stark shift, the latter arising from off-resonant couplings and scaling with \( \Omega(t)^2 \). The ions’ collective motion is primarily governed by the center-of-mass mode, whose frequency sets the detuning, \( \delta \), of the motional sidebands from the carrier.\\
The laser-ion interaction, under the rotating wave approximation, leads to the effective Hamiltonian:
\begin{align}
\hat{H}&=-\eta \Omega(t)\left(a^{\dagger} e^{i \delta t}+a e^{-i \delta t}\right)\notag\\
&\left[S_y \cos (\varphi+\Lambda(t))+S_x \sin (\varphi+\Lambda(t))\right]
\label{eq:effective-hamiltonian-ms}
\end{align}
where \( \eta \) is the Lamb-Dicke parameter, \( a^\dagger \) and \( a \) are phonon ladder operators of the center-of-mass mode, \( S_x \) and \( S_y \) are collective spin operators, \( \varphi \) is the phase of the laser tones, and \( \Lambda(t) \) accounts for accumulated phase errors due to center-line detuning.\\
This configuration enables the realization of entangling Mølmer-Sørensen gates with high fidelity, provided the control parameters are properly calibrated.

\paragraph{Key operational parameters}
An unavoidable time-dependent AC Stark shift introduces deviations from the ideal conditions required for a perfect MS gate. Consequently, the calibration process aims to identify and optimize the control parameters that mitigate the effects of the AC Stark shift, thereby restoring the system to the gate's ideal operating conditions.\\
The control parameters to be optimized are represented by the set
$$
\Theta_c=\left(t_g, f_{c l}, f_{s b}, \phi\right)
$$
which includes the gate duration $t_g$, the mean laser frequency $f_{c l}$, the frequency detuning $f_{s b}$, and the phase difference $\phi$ shared by consecutive gates. Specifically, these parameters are the gate time, the mean frequency of the bichromatic tones, the frequency difference between the tones, and the relative phase between consecutive gate operations.\\
The process combines the theoretical system model with experimental measurements to achieve optimal calibration. This iterative approach updates the control parameters step by step, refining them until the system's behavior aligns with the idealized physics of the MS gate.
We employ an iterative workflow based on Bayesian inference to restore the system to ideal physics. 

\paragraph{Calibration process}
The  \texttt{qruise-toolset} employs a particle filter to approximate the probability distributions of model parameters. This filter represents the parameter space as an ensemble of weighted particles, each corresponding to a possible set of parameter values at a given iteration.\\
The filter is initialized with a Gaussian distribution, characterized by mean values and standard deviations based on prior expectations of the parameters. During each iteration, the particle weights are updated according to insights from the digital twin, and the updated ensemble is used to compute new expected values and variances for the parameters. A particle filtering algorithm based on a Monte Carlo resampling strategy is employed to explore the parameter space efficiently while maintaining a representative distribution. The control parameters are dynamically updated based on the newly inferred model parameters.\\
Measurement settings for subsequent iterations are chosen using calibration strategies designed to minimize the uncertainty in the parameters quickly and effectively (\autoref{app:sensitivity-analysis}). One such strategy is variance minimization, which prioritizes measurement settings that reduce the variances of the particle distribution.\\
Stopping criteria determine when the calibration process has converged, ensuring that the system has reached the desired level of accuracy.

\begin{figure}[!ht]
    \centering
    \begin{subfigure}[h]{0.8\linewidth}
        \centering
        \includegraphics[width=\linewidth]{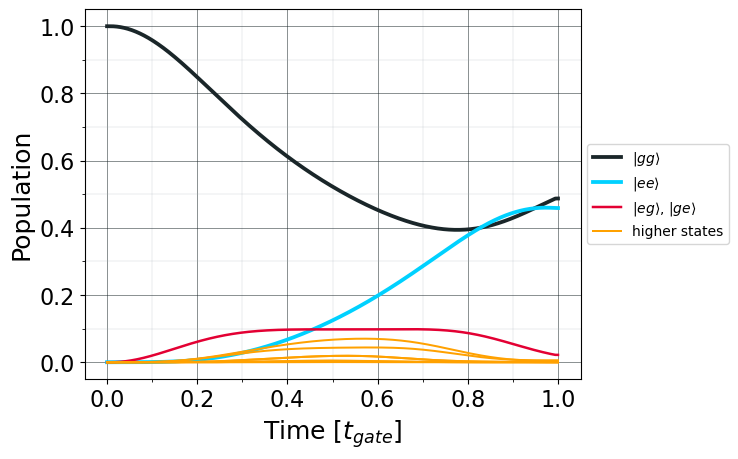}
        \caption{Dynamics before calibration.}
        \label{fig:dynamics_before_calibration}
    \end{subfigure}
    \begin{subfigure}[h]{0.8\linewidth}
        \centering
        \includegraphics[width=\linewidth]{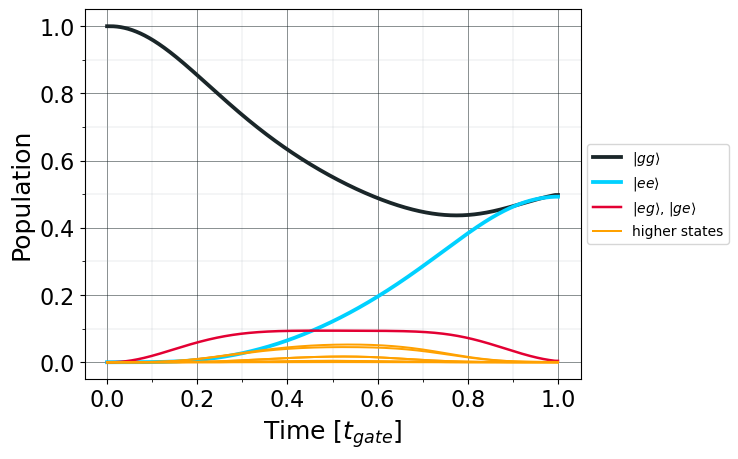}
        \caption{Dynamics after calibration.}
        \label{fig:dynamics_after_calibration}
    \end{subfigure}
    \caption{}
    \label{fig:BayesianCalibrationResults}
\end{figure}
\paragraph{Results}
\autoref{fig:BayesianCalibrationResults} illustrates the gate performance before and after applying the Bayesian calibration algorithm. (a) shows the performance of a non-optimized gate, while (b) depicts the improved performance after calibration. In the calibrated case, the system evolves to achieve a maximally entangled qubit state at the final time, demonstrating the effectiveness of the algorithm. This result highlights the capability of the Bayesian calibration process to restore ideal physical dynamics through the iterative feedback loop between experimental measurements and the digital twin model.\\
In conclusion, Bayesian calibration represents a significant advance for experimental quantum computing, offering a robust framework for refining critical system parameters. Its application to trapped-ion systems enhances efficiency and accuracy, addressing key challenges in maintaining qubit fidelity and coherence. Future research could focus on scaling these methods for larger systems and integrating them with machine learning techniques to further automate the calibration process.

\section{Conclusions}
In this work, we have presented the capabilities of the Qruise software toolset for the modelling, simulation, and optimization of atomic quantum systems. We detailed the modular structure of the modelling framework and highlighted advanced simulation features, including noise generation, quantum optimal control, and benchmarking and calibration tools. These components enable high-fidelity simulation and control of quantum gates under realistic experimental conditions. Through real-world use cases on neutral atom and trapped ion platforms, we demonstrated the toolset’s effectiveness in accurately reproducing system dynamics and significantly enhancing gate performance. Our results confirm that the Qruise toolset is a robust and versatile platform for accelerating the development of high-fidelity quantum operations in state-of-the-art atomic quantum processors. 
\section*{Acknowledgments}
We acknowledge the contributions of several of our former colleagues. Satyanarayana Bade developed the qruise-braket API and helped in the execution \& analysis of experiments on the QuEra Aquila QPU. Yousof Mardoukhi helped the development of the simulation engine and provided insights on noisy simulations of quantum dynamics. Kirsty McGhee reviewed the manuscript and provided helpful advice on content and presentation. We also acknowledge Andrea Alberti and his team at the Max Plank Institute of Quantum Optics, and Robert Zeier and his team at the Forschungzentrum J\"ulich for useful discussions on single- and two-qubit gates on neutral-atom platform. We acknowledge Janine Hilder and her team at neQxt GmbH for valuable inputs regarding the experimental implementations of trapped ion quantum computers.

\bibliographystyle{unsrtnat}
\bibliography{QJ_biblio}

\onecolumn\newpage
\appendix

\section{Quantum process tomography: single- and two-qubit cases}
\label{appendix:qpt}

This appendix expands on the methods used in single- and two-qubit quantum process tomography (QPT) discussed in the main text. We present the mathematical formalism, input-output strategies, and matrix constructions required to reconstruct the process matrix, $\chi$, for both one- and two-qubit channels.

\subsection{Single-qubit process tomography}
\label{app:single-qubit-qpt}

{For single-qubit QPT, the goal is to reconstruct the completely positive trace-preserving map $\mathcal{E}$, which maps an input density operator $\rho$ to its output, using a basis expansion.}

\subsubsection{Operator basis and expansion}
{We used the following operator basis:}
\begin{equation}
\tilde{E}_0 = I, \quad
\tilde{E}_1 = X, \quad
\tilde{E}_2 = -iY, \quad
\tilde{E}_3 = Z.
\end{equation}

Any completely positive trace-preserving map, $\mathcal{E}$, can then be expressed via the $\chi$ matrix as:
\begin{equation}
\mathcal{E}(\rho) = \sum_{m,n=0}^{3} \chi_{mn} \tilde{E}_m \rho \tilde{E}_n^\dagger,
\end{equation}
where $\chi \in \mathbb{C}^{4 \times 4}$ is Hermitian and positive semi-definite.

\subsubsection{Input states and output reconstruction}

We prepare four input states:
\begin{equation}
\begin{aligned}
\rho_1 &= |0\rangle\langle 0|, \\
\rho_2 &= |+\rangle\langle +| = \rho_1 X, \\
\rho_3 &= |-\rangle\langle -| = X \rho_1, \\
\rho_4 &= |1\rangle\langle 1| = X \rho_1 X.
\end{aligned}
\end{equation}

The corresponding outputs under $\mathcal{E}$ are:
\begin{equation}
\begin{aligned}
\rho_1^{\prime} &= \mathcal{E}(|0\rangle\langle 0|), \\
\rho_4^{\prime} &= \mathcal{E}(|1\rangle\langle 1|), \\
\rho_2^{\prime} &= \mathcal{E}(|+\rangle\langle +|) - i \mathcal{E}(|-\rangle\langle -|) - \frac{1 - i}{2}(\rho_1^{\prime} + \rho_4^{\prime}), \\
\rho_3^{\prime} &= \mathcal{E}(|+\rangle\langle +|) + i \mathcal{E}(|-\rangle\langle -|) - \frac{1 + i}{2}(\rho_1^{\prime} + \rho_4^{\prime}).
\end{aligned}
\end{equation}

These are arranged in a $2 \times 2$ block matrix $R$:
\begin{equation}
R =
\begin{bmatrix}
\rho_1^{\prime} & \rho_2^{\prime} \\
\rho_3^{\prime} & \rho_4^{\prime}
\end{bmatrix}.
\end{equation}

\subsubsection{Process matrix reconstruction}

We define the transformation matrix, $\Lambda$:
\begin{equation}
\Lambda = \frac{1}{2}
\begin{bmatrix}
I & X \\
X & -I
\end{bmatrix}.
\end{equation}

Then the $\chi$ matrix is obtained via:
\begin{equation}
\chi = \Lambda R \Lambda.
\end{equation}

\subsubsection{Kraus operator extraction}

The Kraus operators are obtained by diagonalizing $\chi$:
\begin{equation}
\chi_{mn} = \sum_k U_{mk} d_k U_{nk}^*,
\end{equation}
\begin{equation}
A_k = \sqrt{d_k} \sum_j U_{jk} \tilde{E}_j,
\end{equation}
where $U$ is unitary and $d_k$ are eigenvalues of $\chi$.

\subsection{Two-qubit process tomography}
\label{app:two-qubit-qpt}
Two-qubit QPT extends this method using the 16-element operator basis $\{\tilde{E}_i = \sigma_\alpha \otimes \sigma_\beta\}$, with $\sigma_0 = I$, $\sigma_1 = X$, $\sigma_2 = -iY$, $\sigma_3 = Z$. The corresponding process matrix, $\chi$, is $16 \times 16$, with up to 240 real parameters in the trace-preserving case. 
Sixteen linearly independent input states are constructed from tensor products of $\{|0\rangle, |1\rangle, |+\rangle, |+i\rangle\}$. The corresponding output states are arranged in a $4 \times 4$ block matrix and transformed using $\Lambda^{(2)} = \Lambda \otimes \Lambda$ to reconstruct $\chi$.\\
As a general remark, while two-qubit QPT provides detailed characterizations of entangling gates such as CNOT, CZ, or iSWAP, its exponential resource scaling limits practical usage. For larger systems, compressed sensing, randomized benchmarking, and gate set tomography offer scalable alternatives.\\

\subsubsection{Operator basis and expansion}

The basis consists of tensor products:
\begin{equation}
\tilde{E}_i = \sigma_\alpha \otimes \sigma_\beta, \quad \text{for } \alpha, \beta \in \{0,1,2,3\},
\end{equation}
with:
\begin{equation}
\sigma_0 = I, \quad \sigma_1 = X, \quad \sigma_2 = -iY, \quad \sigma_3 = Z.
\end{equation}

Then the quantum process is expanded as:
\begin{equation}
\mathcal{E}(\rho) = \sum_{m,n=0}^{15} \chi_{mn} \tilde{E}_m \rho \tilde{E}_n^\dagger.
\end{equation}

\subsubsection{Input states and tomography}

We used 16 tensor-product input states:
\begin{equation}
\rho_{ij} = \rho_i \otimes \rho_j, \quad \rho_i \in \{ |0\rangle\langle 0|, |1\rangle\langle 1|, |+\rangle\langle +|, |+i\rangle\langle +i| \}.
\end{equation}

Apply the process, $\mathcal{E}$, and reconstruct the output density matrices, $\rho_{ij}'$, via two-qubit state tomography.

\subsubsection{Block matrix and $\chi$ reconstruction}

Arrange the output states into a $4 \times 4$ block matrix $R$, with each entry being a $4 \times 4$ matrix:
\begin{equation}
R =
\begin{bmatrix}
\rho_{00}' & \rho_{01}' & \rho_{02}' & \rho_{03}' \\
\rho_{10}' & \rho_{11}' & \rho_{12}' & \rho_{13}' \\
\rho_{20}' & \rho_{21}' & \rho_{22}' & \rho_{23}' \\
\rho_{30}' & \rho_{31}' & \rho_{32}' & \rho_{33}'
\end{bmatrix}.
\end{equation}

Define the two-qubit transformation:
\begin{equation}
\Lambda^{(2)} = \Lambda \otimes \Lambda.
\end{equation}

Then compute:
\begin{equation}
\chi = \Lambda^{(2)} R \Lambda^{(2)}.
\end{equation}

\subsubsection{Kraus operators}

Extract Kraus operators as in the single-qubit case:
\begin{equation}
\chi = U D U^\dagger, \quad A_k = \sqrt{d_k} \sum_j U_{jk} \tilde{E}_j,
\end{equation}
where $D = \operatorname{diag}(d_0, \dots, d_{15})$.

\subsection{Resource scaling and practical considerations}

While single-qubit QPT is experimentally efficient, two-qubit QPT scales exponentially with the number of qubits. In particular:
\begin{itemize}
    \item Single-qubit QPT requires 4 input states and 4 tomographic measurements.
    \item Two-qubit QPT requires 16 input states and 16 two-qubit state tomography routines.
\end{itemize}

Due to this exponential cost, alternative scalable methods such as randomized benchmarking and gate set tomography are preferred for larger systems, though QPT remains the gold standard for complete gate characterization at small scale.
\section{Closed and open quantum systems}
\label{app:closed-open-quantum-systems}
{
A \textit{closed} quantum system evolves in isolation, without any exchange of energy or information with the environment. Its evolution is governed by the Schr\"odinger equation:
\begin{align*}
    i\hbar \frac{d \ket{\psi}}{dt} = H \ket{\psi},
\end{align*}
where $H$ is the system Hamiltonian and $\ket{\psi}$ is the state vector of the system.
Equivalently, in terms of the density matrix, $\rho(t)$, the system evolution can be described by the von Neumann equation:
\begin{align*}
    i\hbar \frac{d \rho}{dt} = [H, \rho],
\end{align*}
where $[A, B] = AB - BA$ denotes the commutator operator for two operators $A$ and $B$.}

In practice, all quantum systems interact with their environment to some degree, making them \textit{open} quantum systems. These interactions lead to decoherence and dissipation, and are described by the Lindblad master equation:
\begin{align*}
    \frac{d\rho}{dt} = -\frac{i}{\hbar}[H, \rho] + \sum_j \left( 2 L_j \rho L_j^\dagger - \left\{ L_j^\dagger L_j, \rho \right\} \right),
\end{align*}
where $\{A, B\} = AB + BA$ is the anticommutator, and the $L_j$ are Lindblad operators representing different dissipative channels. In contrast to the closed case, the dynamics here are non-unitary.\\
The Lindblad formalism can also be used in a stochastic framework through the method of quantum trajectories, where each trajectory evolves under an effective non-Hermitian Hamiltonian:
\begin{align*}
    H_\text{eff} = H - \frac{i}{2} \sum_k L_k^\dagger L_k.
\end{align*}
Averaging over many such trajectories recovers the full Lindblad evolution. The \texttt{qruise} software supports both deterministic and trajectory-based approaches to simulate open system dynamics.

\section{Benchmark techniques}
This appendix describes two complementary protocols for characterizing the performance of quantum processors. Randomized benchmarking (RB) estimates the average error of a gate set while reducing sensitivity to state-preparation-and-measurement (SPAM) errors. Quantum volume (QV) evaluates the largest class of random square circuits that a processor can execute reliably, thereby combining the effects of gate errors, qubit connectivity, and circuit parallelism. Whereas RB provides a gate-level characterization, QV provides a circuit-level benchmark of the processor's usable computational capacity.

\label{app:benchmark-techniques}
\subsection{Randomized benchmarking}
{Randomized benchmarking (RB) estimates the average fidelity of a gate set by applying sequences of randomly selected Clifford operations. For each random sequence, a final inversion Clifford is appended such that, in the absence of errors, the complete sequence implements the identity operation and returns the initial state to itself. The error per Clifford is then inferred from the decay of the measured survival probability as the sequence length increases.}

The \texttt{qruise-toolset} integrates RB into its simulation and hardware interfaces via Qiskit, supporting the evaluation of user-defined Hamiltonians. This enables a platform-independent fidelity metric robust to SPAM errors.

\subsection{Quantum Volume}
Quantum volume (QV) is a hardware-independent benchmark of a quantum processor's usable circuit capacity. It incorporates the combined effects of gate fidelity, qubit connectivity, and circuit parallelism. The QV is defined as $2^d$, where $d$ is the largest square-circuit dimension for which the processor successfully executes representative random circuits of width and depth $d$. For each circuit, the ideal simulation identifies the output bit strings with probabilities above the median; these are called heavy outputs. The processor passes at dimension $d$ when its measured probability of producing heavy outputs exceeds the protocol threshold.
Qruise implements the quantum-volume (QV) protocol in four stages. For a candidate circuit dimension $d$, the protocol generates $n_c$ random square circuits, each with width $d$ and depth $d$. An ideal simulation of each circuit identifies its \emph{heavy outputs}, defined as the bit strings whose ideal output probabilities are greater than the median probability of
that circuit's output distribution. Each compiled circuit is executed with $n_s$ shots, and the heavy-output probability is evaluated for every circuit.

\begin{enumerate}
    \item \textbf{Circuit generation:} Random square circuits of width and depth $d$ are generated using Qiskit's \texttt{quantum\_volume} module.
    \item \textbf{Transpilation:} Each abstract circuit is transpiled to the native gate set and connectivity constraints of the modeled backend.
    \item \textbf{Pulse compilation:} The transpiled circuit is mapped to analog control pulses using Qruise's modular signal chain.
    \item \textbf{Execution and analysis:} Each compiled circuit is executed with $n_s$ shots. For circuit $i$, let $h_i$ denote the number of measured heavy outputs. Its heavy-output probability is
    \begin{equation}
        p_{h,i} = \frac{h_i}{n_s}.
    \end{equation}
    The mean heavy-output probability over the $n_c$ circuits is
    \begin{equation}
        \overline{p}_h = \frac{1}{n_c}\sum_{i=1}^{n_c} p_{h,i}.
    \end{equation}
    Its standard error is estimated from the circuit-to-circuit sample standard deviation,
    \begin{equation}
        \operatorname{SE}(\overline{p}_h)
        =
        \sqrt{
            \frac{1}{n_c(n_c-1)}
            \sum_{i=1}^{n_c}
            \left(p_{h,i}-\overline{p}_h\right)^2
        }.
    \end{equation}
    The device passes at circuit dimension $d$ when the approximate two-standard-deviation lower confidence bound exceeds $2/3$:
    \begin{equation}
        \overline{p}_h
        -
        2\operatorname{SE}(\overline{p}_h)
        >
        \frac{2}{3}.
    \end{equation}
\end{enumerate}

The quantum volume is reported as $2^d$, where $d$ is the largest circuit dimension that satisfies this success criterion.
Here $n_c$ is the number of random circuits at dimension $d$, $n_s$ is the number of shots per circuit, $h_i$ is the number of heavy-output observations for circuit $i$, $p_{h,i}$ is the heavy-output probability of circuit $i$, $\overline{p}_h$ is the mean heavy-output probability, and $2\operatorname{SE}(\overline{p}_h)$ is the approximate 95\% lower-confidence margin.

\begin{figure}[ht!]
    \centering
    \includegraphics[width=0.5\linewidth]{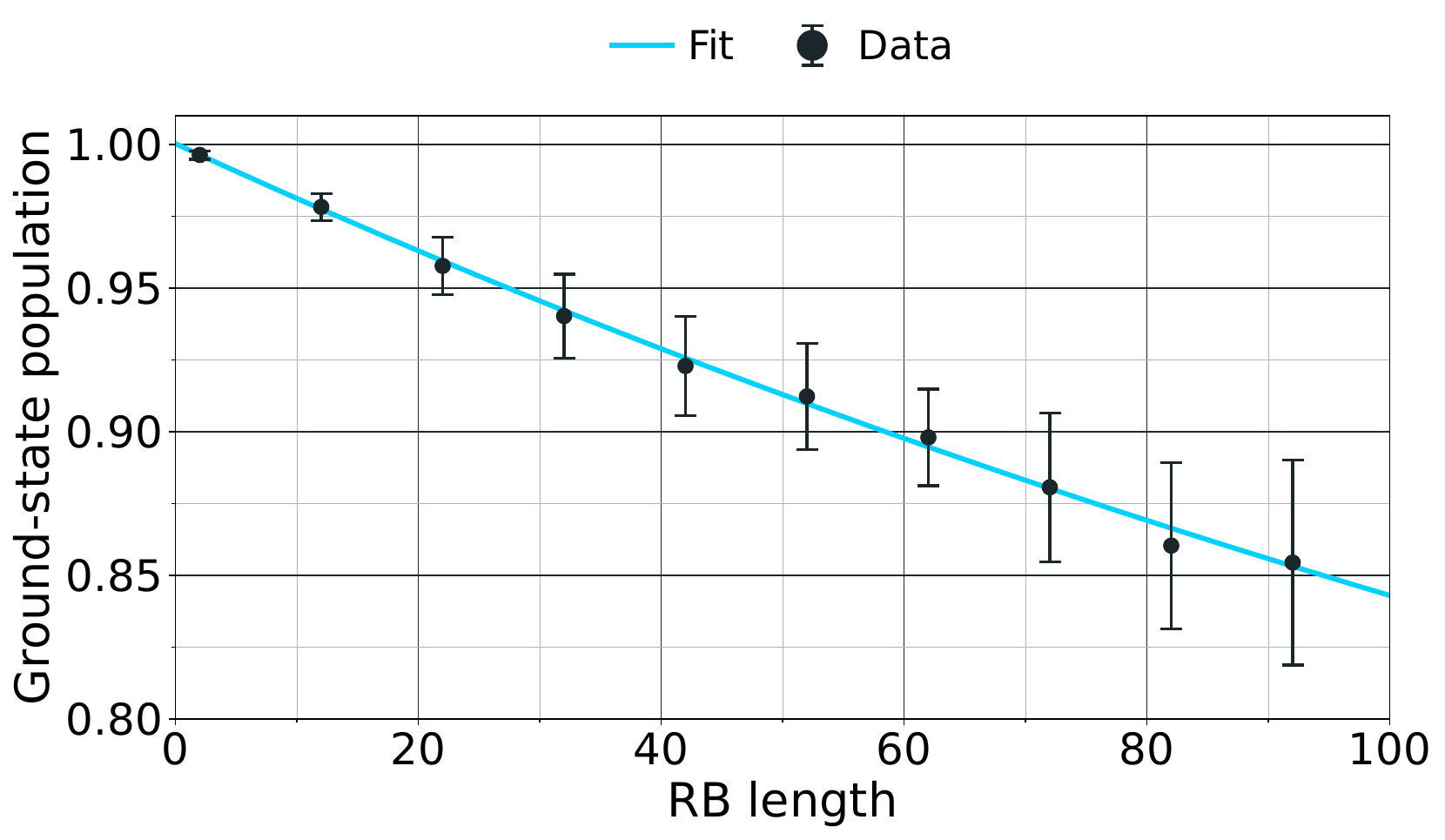}
    \caption{Simulated single-qubit randomized benchmarking. The ground-state population is measured after Clifford sequences of increasing length (markers: mean over 40 random sequences, each averaged over three drive settings; error bars: standard deviation over the sequences) and fitted with an exponential decay $A p^n + b$ (solid line). The fitted error per Clifford is $1.11(30)\times10^{-3}$.}
    \label{fig:randomized-benchmarking}
\end{figure}
\begin{figure}
    \centering
    \includegraphics[width=0.6\linewidth]{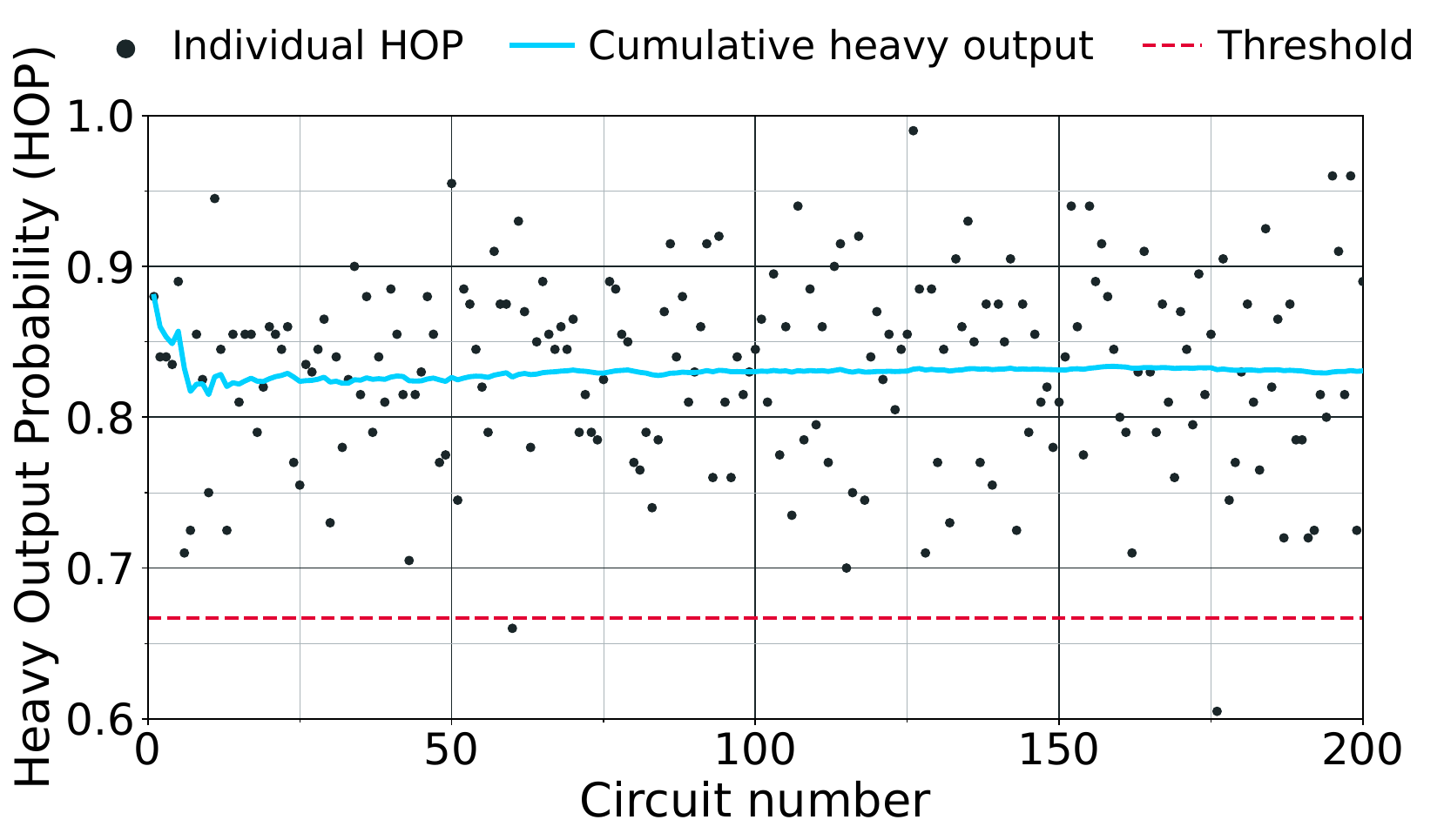}
    \caption{Quantum-volume benchmark of the simulated four-qubit Rydberg-atom register. Each marker is the heavy-output probability (HOP) of one random square circuit of width and depth four, evaluated from 200 shots per circuit; heavy outputs are the ideal bit strings whose simulated probabilities exceed the median. The solid line is the cumulative mean HOP over the 200 circuits, and the dashed line marks the $2/3$ pass threshold. The mean HOP reaches $83.1\%$, above the threshold, so the processor passes at dimension $d=4$, corresponding to a quantum volume of $2^4=16$.}
    \label{fig:quantum_volume_example}
\end{figure}
\section{Recoil-free protocol optimization}
\label{app:recoil-free-protocol}
The quantum process fidelity is given by:
\begin{align*}
\langle\mathcal{F}\rangle &= \frac{1 + n \sum_{k=0}^3 \chi_k \left| \operatorname{Tr} \left( U_{\text{target}}^{\dagger} E_k \right) / n \right|^2 }{1 + n} \\
& \approx \frac{1 + n \chi_0 \left| \operatorname{Tr} \left( U_{\text{target}}^{\dagger} E_0 \right) / n \right|^2 }{1 + n}
\end{align*}
where \( n = 2 \) is the dimension of the qubit Hilbert space, \( \chi_k \) are the elements of the process matrix, \( E_k \) form an operator basis, and \( U_{\text{target}}\) is the target unitary. The approximation assumes that most of the process is captured by the \( k = 0 \) component.
We proceed as follows:
\begin{enumerate}
    \item Definition of the initial states
    \item Evolution of the initial states through the quantum process
    \item Calculation of the reduced density matrices
    \item Averaging over the Boltzmann distribution
    \item Calculation of the $\chi$ process matrix
    \item Calculation of the Kraus operators
\end{enumerate}
\subsection{Initial states}
The set of initial states is a tensor product between the qubit state and the motional state. In total, we have 4N initial states, where N is the Hilbert space dimension of the motional state.
\begin{equation}
\begin{aligned}
& \left|\psi_1\right\rangle(n)=|0\rangle|n\rangle \\
& \left|\psi_2\right\rangle(n)=|1\rangle|n\rangle \\
& \left|\psi_3\right\rangle(n)=|+\rangle|n\rangle=\frac{1}{\sqrt{2}}(|0\rangle+|1\rangle)|n\rangle \\
& \left|\psi_4\right\rangle(n)=|-\rangle|n\rangle=\frac{1}{\sqrt{2}}(|0\rangle+i|1\rangle)|n\rangle
\end{aligned}
\end{equation}
\subsection{Evolution of the initial states through the quantum process}
We solve the Schr\"odinger equation associated with the quantum process
$$
\frac{\partial}{\partial t} \mathcal{U}(t)=-i H \mathcal{U}(t)
$$

where

$$
\mathcal{U}(t)=\mathcal{T} e^{-i \int_0^T d t H(t)}
$$
with $\mathcal{T}$ denoting the time-ordering operator
and propagate the previous defined initial states:
$$
\left|\psi_{f, i}\right\rangle(n)=\mathcal{U}(T)\left|\psi_i\right\rangle(n)
$$
\subsection{Calculation of the reduced density matrices}
For each final state, we calculate the reduced density matrix, tracing out the motional state:

$$
\rho_{q, i}^n=T_{r_n}\left(\left|\psi_{f, i}^n\right\rangle\left\langle\psi_{f, i}^n\right|\right)
$$

where $\quad\left|\psi_{f,i}^n\right\rangle=\left|\psi_{f, i}\right\rangle(n)$.
\subsection{Averaging over the Boltzmann distribution}
We calculate the mixed qubit state by averaging over the possible initial motional states, weighted according to their thermal populations.

$$
\rho_{q, i}=\sum_n p_n \rho_{q, i}^n
$$

where

$$
p_n=\frac{e^{n \ln \left(1-p_0\right)}}{\sum_k e^{k \ln \left(1-p_0\right)}}
$$
are the weights defined by the Boltzmann distribution.
\subsection{Calculation of the chi process matrix}
We calculate the following basis states for the quantum process tomography:

$$
\begin{gathered}
\rho_1^{\prime}=\rho_{q, 1} \\
\rho_4^{\prime}=\rho_{q, 2} \\
\rho_2^{\prime}=\rho_{q, 3}-i \rho_{q, 1}-\frac{(1-i)}{2}\left(\rho_1^{\prime}+\rho_4^{\prime}\right) \\
\rho_3^{\prime}=\rho_{q, 3}+i \rho_{q, 4}-\frac{(1+i)}{2}\left(\rho_1^{\prime}+\rho_4^{\prime}\right)
\end{gathered}
$$
and the $\chi$ process matrix
$$
\begin{aligned}
& \chi_{p m}=\Lambda\left[\begin{array}{ll}
\rho_1^{\prime} & \rho_2^{\prime} \\
\rho_3^{\prime} & \rho_4^{\prime}
\end{array}\right]
\end{aligned}
$$
where
$$
\begin{aligned}
& \Lambda=\frac{1}{2}\left[\begin{array}{cc}
\mathcal{I}_2 & \sigma_x \\
\sigma_x & -\mathcal{I}_2
\end{array}\right]
\end{aligned}
$$
\subsection{Calculation of the Kraus operators
}
To derive the Kraus operators, we diagonalize the chi process matrix:

$$
\left(\chi_{p m}\right)_{r s}=\sum_{k, l} U_{r k} \chi_k \delta_{k l} U_{s l}^*
$$

Finally, the Kraus operators are defined as

$$
A_i=\sqrt{\chi_i} \sum_j U_{j i} \tilde{E}_j=\sqrt{\chi_i} E_i
$$

where

$$
\begin{gathered}
\tilde{E}_0=\mathcal{I}_2 \\
\tilde{E}_1=\sigma_x \\
\tilde{E}_2=-i \sigma_y \\
\tilde{E}_3=\sigma_z
\end{gathered}
$$
\subsection{Summary of quantum process tomography
}
We started from a set of initial states for the system (the qubit and motional states) and obtained a quantum process for only the qubit subspace.
This is defined as follows:

$$
\mathcal{E}\left(\rho_q\right)=\sum_{i=0}^3 A_i \rho_q A_i^{\dagger}
$$
where $\rho_q$ is an arbitrary qubit state, and $A_i$ satisfies the condition $\operatorname{Tr}\left(A_i^{\dagger} A_j\right)=\delta_{i j}$.
We can rewrite the expression for the quantum process as

$$
\mathcal{E}\left(\rho_q\right)=\sum_{i=0}^3 \chi_i E_i \rho_q E_i^{\dagger}
$$
where $\chi_i$ are the eigenvalues of the quantum process.
They are ordered in decreasing order and sum to 1.

In the limit of the unitary process, the $0^\text{th}$ eigenvalue is equal to 1, and the $0^\text{th}$ Kraus operator is unitary. Physically, this means that the entire quantum process, including the qubit and motional state, leaves the two subsystems uncoupled at the final time, or, in other words, the system is no longer entangled.

\section{Relevant infidelities for two-qubit gate}
\label{app:infidelities-2-qubit-gate}
Here, we define the infidelity for the two-qubit gate.
First, we define the following set of labels and states:\\
The set of the clock qubit state is defined as $$\vec{q} = [q_1, q_2] \in \{[0,0],[1,0],[0,1],[1,1]\}.$$\\
The Fock state for the $i^\text{th}$ motional state is
$$\ket{n_i}=\{\ket{0},\ket{1}, \cdots,\ket{N}\} \forall i \in \{1,2\}.$$\\
For a given pair of qubit states $(q_1,q_2)$ and motional Fock states $(n_1,n_2)$, the initial state of the complete system is
$$\left|\psi_0(q_1,q_2,n_1,n_2)\right\rangle=\ket{q_1}\ket{q_2}\ket{n_1}\ket{n_2}.$$\
The corresponding initial motional state is
$$\ket{\psi_{0,\mathrm{ms}}(n_1,n_2)}=\ket{n_1}\ket{n_2}.$$\
After evolution under the propagator $U(T)$, the final state is
$$\left|\psi_f(q_1,q_2,n_1,n_2)\right\rangle=U(T)\left|\psi_0(q_1,q_2,n_1,n_2)\right\rangle.$$\
Tracing out the motional degrees of freedom gives the reduced qubit state
\begin{equation}
    \rho_{n_1,n_2}(q_1,q_2)=\operatorname{Tr}_{\mathrm{motion}}\!\left[\ket{\psi_f(q_1,q_2,n_1,n_2)}\bra{\psi_f(q_1,q_2,n_1,n_2)}\right].
\end{equation}

\subsection{Entanglement infidelity}
We can estimate the entanglement between the motional states and the qubits by the following fidelity
\begin{equation*}
    j_{\text {ent }}\left(q_1, q_2, n_1, n_2\right)=1-\sqrt{\left\langle\psi_{0, m s}\right| \rho_{n_1, n_2}\left|\psi_{0, m_s}\right\rangle}
\end{equation*}
Averaging over different motional level realizations with probabilities according to the boltzmann distribution we have
\begin{equation*}
    J_{\text {ent}}=\sum_{n_1, n_2} p_{n_1} p_{n_2}\left(\frac{j_{\text {ent }}\left(1,0, n_1, n_2\right)}{2}+ \frac{j_{\text {ent }}\left(1,1, n_1, n_2\right)}{4}\right)
\end{equation*}

\subsection{Leakage error}
Leakage quantifies the population transferred outside the two-qubit computational subspace. Let $P_{\mathrm{comp}}$ denote the projector onto that subspace, and let $\rho_{\mathrm{int}}(q_1,q_2,n_1,n_2)$ be the final internal-state density matrix after tracing out the motional degrees of freedom. The leakage for a given initial state is then
\begin{equation*}
    j_{\text {leak }}\left(q_1, q_2, n_1, n_2\right)=1-\operatorname{Tr}\!\left[P_{\mathrm{comp}}\,\rho_{\mathrm{int}}(q_1,q_2,n_1,n_2)\right].
\end{equation*}
After averaging over the initial motional Fock states with their Boltzmann probabilities, the total leakage cost is
\begin{equation}
    J_{\text {leak }}=\sum_{n_1,n_2} p_{n_1}p_{n_2}\left[\frac{j_{\text {leak }}(1,0,n_1,n_2)}{4}+\frac{j_{\text {leak }}(1,1,n_1,n_2)}{8}\right].
\end{equation}
\subsection{Phase infidelity}
The infidelity for the phase difference can be estimated via
\begin{equation*}
    j_{\text {phase }}\left(q_1, q_2, n_1, n_2\right)=\operatorname{angle}\left(\left\langle\psi_f \mid \psi_0\right\rangle\right)
\end{equation*}
Averaging over different motional level realizations with probabilities according to the Boltzmann distribution, we have
\begin{equation}
    J_{\text {phase }}=\frac{3}{32} \sum_{n_1, n_2} p_{n_1} p_{n_2}\left(\phi_C\left(n_1, n_2\right)-\pi\right)^2
\end{equation}
with
\begin{equation*}
    \phi_C\left(n_1, n_2\right)=j_{\text {phase }}\left(1,1, n_1, n_2\right)-j_{\text {phase }}\left(1,0, n_1, n_2\right)-j_{\text {phase }}\left(0,1, n_1, n_2\right)
\end{equation*}

\section{M\o lmer-S\o rensen gate optimal parameters}
\label{app:ms-gate-optimal-parameters}
The M\o lmer-S\o rensen (MS) gate is denoted $\operatorname{MS}_{\varphi}(\theta)$, where the upright $\operatorname{MS}$ names the gate itself, the subscript $\varphi$ is the phase of the driving laser tones, and $\theta$ is the entangling angle accumulated during the gate. Its action is
\begin{equation}
\operatorname{MS}_{\varphi}(\theta)=\exp \left(-i \theta S_{\varphi}^2\right),
\label{eq:ms-gate-definition}
\end{equation}
where $S_\varphi$ is the same collective spin operator introduced in \autoref{eq:effective-hamiltonian-ms}, expressed as
\begin{equation}
S_{\varphi}=S_y \cos (\varphi)+S_x \sin (\varphi).
\label{eq:ms-spin-operator}
\end{equation}
The gate is maximally entangling for $\theta=\pi/2$.\\
Assuming a zero accumulated phase, i.e.\ $\Lambda(t)=0$, the evolution operator generated by the Hamiltonian of \autoref{eq:effective-hamiltonian-ms} can be integrated analytically as
\begin{equation}
\hat{U}(t)=D\left[\gamma(t) S_{\varphi}\right] \exp \left[i \theta(t) S_{\varphi}^2\right],
\label{eq:ms-propagator}
\end{equation}
where $D[\alpha]=\exp\left(\alpha a^\dagger - \alpha^* a\right)$ is the displacement operator acting on the shared motional mode, and
\begin{equation}
\gamma(t)=i \eta \int_0^t \Omega\left(t^{\prime}\right) e^{i t^{\prime} \delta} \, d t^{\prime}
\label{eq:ms-gamma}
\end{equation}
is the (generally nonzero) spin-motion coupling accrued during the pulse, while
\begin{equation}
\theta(t)=\eta^2 \operatorname{Im} \int_0^t d t^{\prime} \int_0^{t^{\prime}} d t^{\prime \prime} \, \Omega\left(t^{\prime}\right) \Omega\left(t^{\prime \prime}\right) e^{-i \delta\left(t^{\prime \prime}-t^{\prime}\right)}
\label{eq:ms-theta}
\end{equation}
is the accumulated entangling angle.\\
A perfect MS gate requires the ions' internal and motional states to be disentangled at the end of the gate, i.e.\ $\gamma(t_g)=0$, together with a maximal entangling angle $\theta(t_g)=\pi/2$. For a constant Rabi frequency $\Omega$, these two conditions are satisfied simultaneously by choosing the gate duration and detuning such that
\begin{equation}
\frac{(\eta \Omega)^2 t_g}{\delta}=\frac{\pi}{2}
\label{eq:ms-condition-theta}
\end{equation}
and
\begin{equation}
|\delta|=\frac{2\pi}{t_g}.
\label{eq:ms-condition-loop}
\end{equation}
\autoref{eq:ms-condition-loop} ensures that the motional trajectory in phase space closes exactly one loop over the gate duration, so that $\gamma(t_g)=0$ regardless of the initial motional state, while \autoref{eq:ms-condition-theta} fixes the corresponding entangling angle to $\pi/2$. Using these analytically derived values as the experimental parameters therefore yields a maximally entangling gate between the two qubits.
\section{Bayesian calibration sensitivity analysis}
\label{app:sensitivity-analysis}
\begin{figure}[!htb]
    \centering
    \begin{subfigure}[t]{0.4\linewidth}
        \centering
        \includegraphics[width=\linewidth]{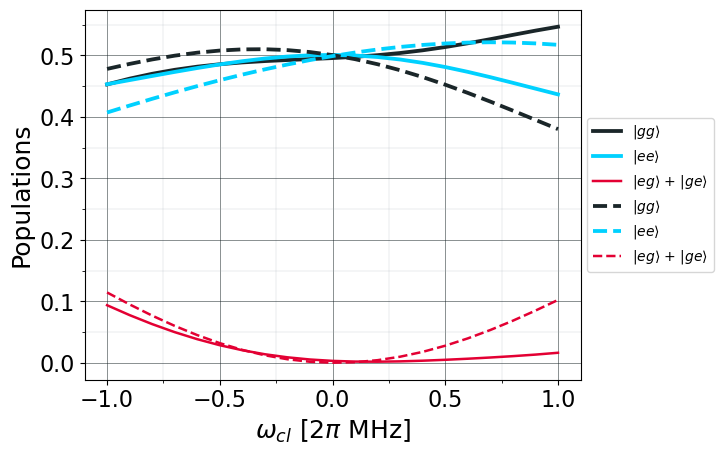}
        \caption{Sensitivity to the center-line frequency.}
    \end{subfigure}
    \begin{subfigure}[t]{0.4\linewidth}
        \centering
        \includegraphics[width=\linewidth]{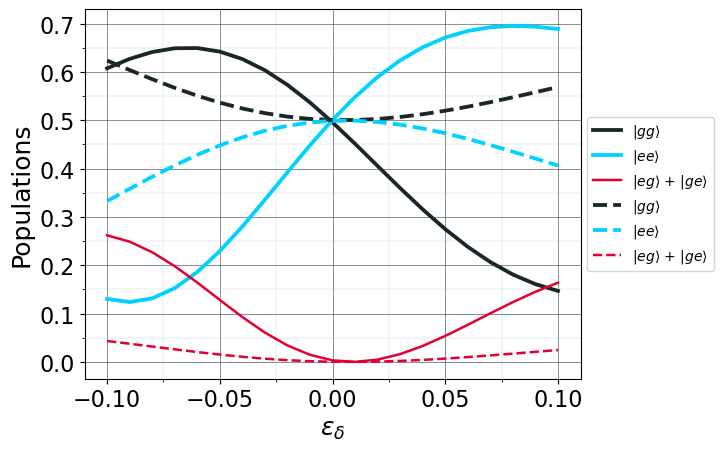}
        \caption{Sensitivity to the sideband detuning.}
    \end{subfigure}
    \begin{subfigure}[t]{0.4\linewidth}
        \centering
        \includegraphics[width=\linewidth]{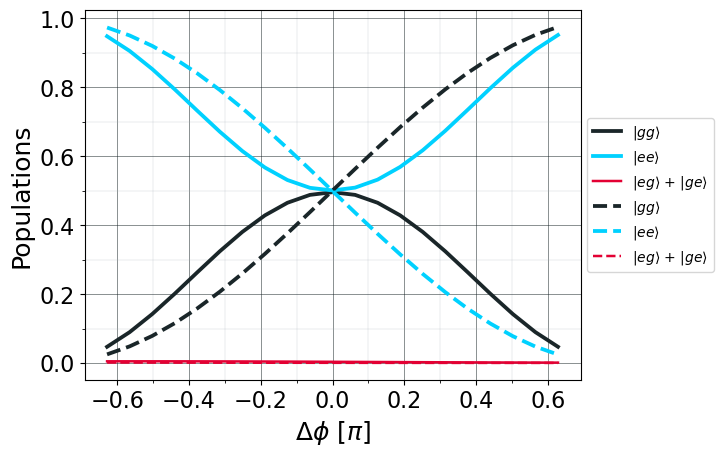}
        \caption{Sensitivity to the gate phase.}
    \end{subfigure}
    \begin{subfigure}[t]{0.4\linewidth}
        \centering
        \includegraphics[width=\linewidth]{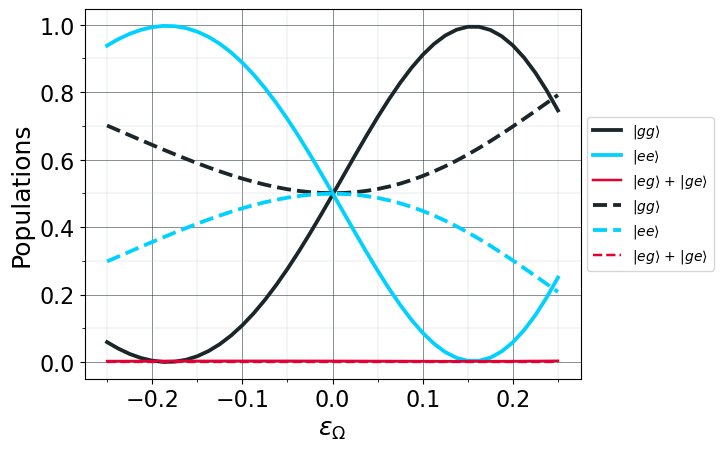}
        \caption{Sensitivity to the Rabi frequency.}
    \end{subfigure}
    \caption{Population sensitivity of two candidate gate sequences, the two-pulse sequence $\mathrm{MS}_0\left(\frac{\pi}{2}\right) \mathrm{MS}_{\frac{\pi}{4}}\left(\frac{\pi}{2}\right)$ (dashed line) and the three-pulse sequence $\mathrm{MS}_0\left(\frac{\pi}{2}\right)\mathrm{MS}_0\left(\frac{\pi}{2}\right)\mathrm{MS}_0\left(\frac{\pi}{2}\right)$ (solid line), to miscalibration of each control parameter in $\Theta_c$.}
    \label{fig:BayesianCalibrationSensitivity}
\end{figure}

We construct a digital twin of the trapped-ion experiment to analyze the system's sensitivity. Using the digital twin, we investigate how the populations evolve across various ranges of the model parameters. 
This analysis enables us to quantify the experiment's sensitivity to parameter miscalibrations, providing insights into the system's robustness. Additionally, it informs decisions about which calibration protocols are most effective, guiding strategies to enhance convergence and optimize the parameter landscape.\\
\autoref{fig:BayesianCalibrationSensitivity} shows, for two candidate gate sequences, how the final-state population responds to a miscalibration of each control parameter $\Theta_c$: the center-line frequency $f_{cl}$ (panel a), the sideband detuning $f_{sb}$ (panel b), the relative gate phase $\phi$ (panel c), and the Rabi frequency $\Omega$ (panel d). Comparing the two curves in each panel reveals which sequence is more robust to that particular miscalibration, and thereby which calibration protocol should be used to correct it.
In \autoref{fig:BayesianCalibrationSensitivity}(c), the dashed two-pulse sequence is first-order insensitive to phase miscalibrations: the outcome probability has a local minimum at zero miscalibration, so the population barely changes for small phase errors, and its symmetry about zero also means the \emph{sign} of the miscalibration cannot be determined from the population alone. The same sequence is, however, first-order sensitive to Rabi-frequency miscalibrations, as shown by the steep, antisymmetric response of the dashed curve in panel (d).
For the solid three-pulse sequence, this pattern is reversed: it is first-order sensitive to phase miscalibrations but first-order insensitive to Rabi-frequency miscalibrations. In other words, a calibration protocol targeting the gate phase should use the three-pulse sequence, while one targeting the Rabi frequency should use the two-pulse sequence. The original text further associated this same trade-off with panels (a) and (b), stating that the qualitative sensitivity to the center-line frequency mirrors the sensitivity to phase, and the sensitivity to the sideband detuning mirrors the sensitivity to the Rabi frequency.
\end{document}